\documentclass[%
reprint,
 amsmath,amssymb,
 aps,
prb
]{revtex4-2}
\usepackage{graphicx}
\usepackage{dcolumn}
\usepackage{bm}
\usepackage{comment}
\usepackage{amsthm}
\usepackage{amsmath}
\usepackage{mathtools}
\newtheorem{theoremP}{PTM Rule}
\newtheorem{theoremPP}{Poly-PTM Rule}
\begin{document}

\preprint{APS/123-QED}

\newcommand{\alp}{$\alpha$}
\newcommand{\bet}{$\beta$}
\newcommand{\alpu}[1]{$\alpha_{#1}$}
\newcommand{\betu}[1]{$\beta_{#1}$}
\newcommand{\alpptm}{$\alpha$-PTM}
\newcommand{\betptm}{$\beta$-PTM}
\newcommand{\alpptms}{$\alpha$-PTMs}
\newcommand{\betptms}{$\beta$-PTMs}
\newcommand{\alpbetptm}{$\alpha$($\beta$)-PTM}
\newcommand{\betalpptm}{$\beta$($\alpha$)-PTM}
\newcommand{\alpbetptms}{$\alpha$($\beta$)-PTMs}
\newcommand{\betalpptms}{$\beta$($\alpha$)-PTMs}
\newcommand{\sqthree}{$\sqrt{3} \times \! \sqrt{3}$}
\newcommand{\mrm}[1]{\mathrm{#1}}
\newcommand{\trans}[1]{\nolinebreak{^t\!{#1}}}
%

\title{Zero-energy modes in super-chiral nanographene networks of phenalenyl-tessellation molecules}

\author{Naoki Morishita$^{1}$}
\email{morishi@artemis-mp.jp}
\author{Koichi Kusakabe$^1$}%
 \email{kabe@mp.es.osaka-u.ac.jp}
\affiliation{%
 $^1$Graduate School of Engineering Science, Osaka University, Toyonaka, Osaka 560-8531, Japan\\
}%

\date{\today}

\begin{abstract}
We have derived a general rule for the appearance of zero-energy modes in super-chiral defective nanographene.
This so-called ``super-zero-sum rule'' defines the appearance of zero modes in a new class of materials, which we call polymerized phenalenyl-tessellation molecules (poly-PTMs).
Through theoretical modeling of the electronic states in these molecular forms, 
we provide concrete solutions for achieving the quantum-spin systems needed in quantum-information devices.
The two-dimensional graph of electronic $\pi$-orbitals in the poly-PTM possesses a number of localized zero modes 
equivalent to that of vacancies in PTMs.
In addition to the modes confined to each PTM, 
another type of zero mode may appear according to the super-zero-sum rule supported by super-chirality. 
Since the magnetic interactions among quantum spins in the zero modes are determined by how they appear (which is governed by the super-zero-sum rule), 
our rule is indispensable for designing quantum-information devices using electron zero modes in poly-aromatic hydrocarbons and defective graphene with vacancies.

\end{abstract}

\maketitle



\section{\label{Intro}Introduction}
The unconventional electronic structure of graphene\cite{Novoselov666,Geim2007} 
has attracted a great interest from researchers 
because of the emergence of the massless Dirac fermion\cite{Novoselov2005, RevModPhys.81.109}.
One unique characters of the Dirac fermion system is the appearance of zero-energy modes, 
which are typically found at the edges\cite{doi:10.1143/JPSJ.65.1920, PhysRevB.54.17954, PhysRevLett.89.077002, KLUSEK2000508, PhysRevB.71.193406, NIIMI200543, PhysRevB.73.045124, PhysRevB.87.115427, Fujii2014, Ziatdinov2017, doi:10.7566/JPSJ.87.084706} 
and the vacancies\cite{PhysRevLett.96.036801,PhysRevLett.98.259902,PhysRevB.77.115109,PhysRevLett.93.187202,PhysRevB.89.155405} of graphene.
Because the zero modes at Fermi energy are naturally half-filled 
(with one electron occupying each mode at the charge neutral point), 
and since unpaired electrons can become magnetically active, 
the physics and chemistry arising from the zero modes have been highlighted, 
creating a vast field in science\cite{2014,2019}. 
Indeed, since a single-electron spin may appear in each zero mode orbital because of the electron correlation effect, 
electron spins can create a Heisenberg spin system as a whole, 
such that useful magnetic functions can appear in graphene with zero modes.

Zero modes have potential uses as quantum bits, 
since a single-electron spin may appear in the orbital 
and unpaired electrons may show a nontrivial correlation because of the strong correlation effects. 
There are several types of zero modes supporting localized electrons, 
which coexist with Dirac electrons.
Therefore, it is important to find a general rule for controlling the appearance of zero modes.
One example is Shima and Aoki's classification of the appearance of gapless linear dispersion and flat bands\cite{PhysRevLett.71.4389}.
The appearance of the flat band at the zero-energy indicates that 
a huge number of zero modes exist in their honeycomb networks.
However, a rule based on crystal lattice symmetry is not  
applicable to general molecular structures and networks with a lower symmetry than 
that of the periodic honeycomb lattice.

Singular enhancement in the density of state around the Dirac point 
has been discussed as an effect of a chiral defect (e.g. atom vacancies),
which are often assumed to be randomly distributed\cite{PhysRevLett.115.106601}.
The related density of states in a model with structural defects 
have been discussed by specifying several examples\cite{PhysRevB.101.235116}; 
however, without finding a control rule for certifying zero-energy eigenstates in carbon networks, 
this discussion is inadequate for purposes such as designing devices for quantum computers, 
for which more precise control of the electron spins would be required.


For a definite series of nanographene networks, 
the authors reported a concrete example of a rule to obtain the zero modes, 
even for a system with no sublattice imbalance\cite{doi:10.7566/JPSJ.85.084703, doi:10.7566/JPSJ.86.034802}.
Although this work was followed by consideration of networks with decorated edges\cite{PhysRevB.94.064204},
two types of zero modes are defined for our vacancy-centered systems, 
successfully realizing an embedded localized zero mode as an eigenstate.
It was discussed that a well-defined boundary condition of a supposed zero-energy eigenstate becomes artificial,
which is given only by a mode expansion fomula\cite{PhysRevB.102.075109}.
Our systems, however, provided a means of defining the true Dirichlet boundary condition for 
the zero mode wavefunction $\psi(\bm{r})$ as $\psi(\bm{r})|_\mrm{boundary}=0$\cite{doi:10.7566/JPSJ.85.084703}.
We have also recently discovered a class of graphene nanomolecular structures 
that we call phenalenyl-tessellation molecules (PTMs)\cite{doi:10.7566/JPSJ.88.124707}.
When a PTM contains atomic defects, two types of special non-bonding molecular orbitals appear at the zero-energy level;
one is a vacancy-originated localized zero mode, and the other is an extended Dirac zero mode.

This paper reports a general rule for the appearance of the Dirac and localized zero modes on a graph of 
$\pi$-electron systems of a class of PTM-based poly-aromatic hydrocarbons. 
Our new rule generally holds in molecular and periodic systems, even with low spatial symmetry.
However, we discovered the relevance of generalized chiral symmetry in a super-structure (or super-chiral symmetry) in super networks of PTMs.

A quantum spin in the zero mode has the potential to be used as an easy-to-access quantum bit.
Furthermore, plenty of zero modes can be created in a two-dimensional $\pi$-orbital lattice of graphene structure.
If the distribution of spins and the connections among them can be engineered to yield a desired spin system, 
these materials may be useful as units for quantum computation.
Each unit that possesses a unique unitary transformation in its time-evolution step 
may work as a computational element.
Our method of producing zero modes is flexible enough to design quantum devices using graphene nanostructures; 
therefore, our rule is indispensable for the design of quantum-information devices.



\section{Method}
\subsection{Construction of poly-PTM structures}
\begin{figure}[b]
\includegraphics[width=86mm]{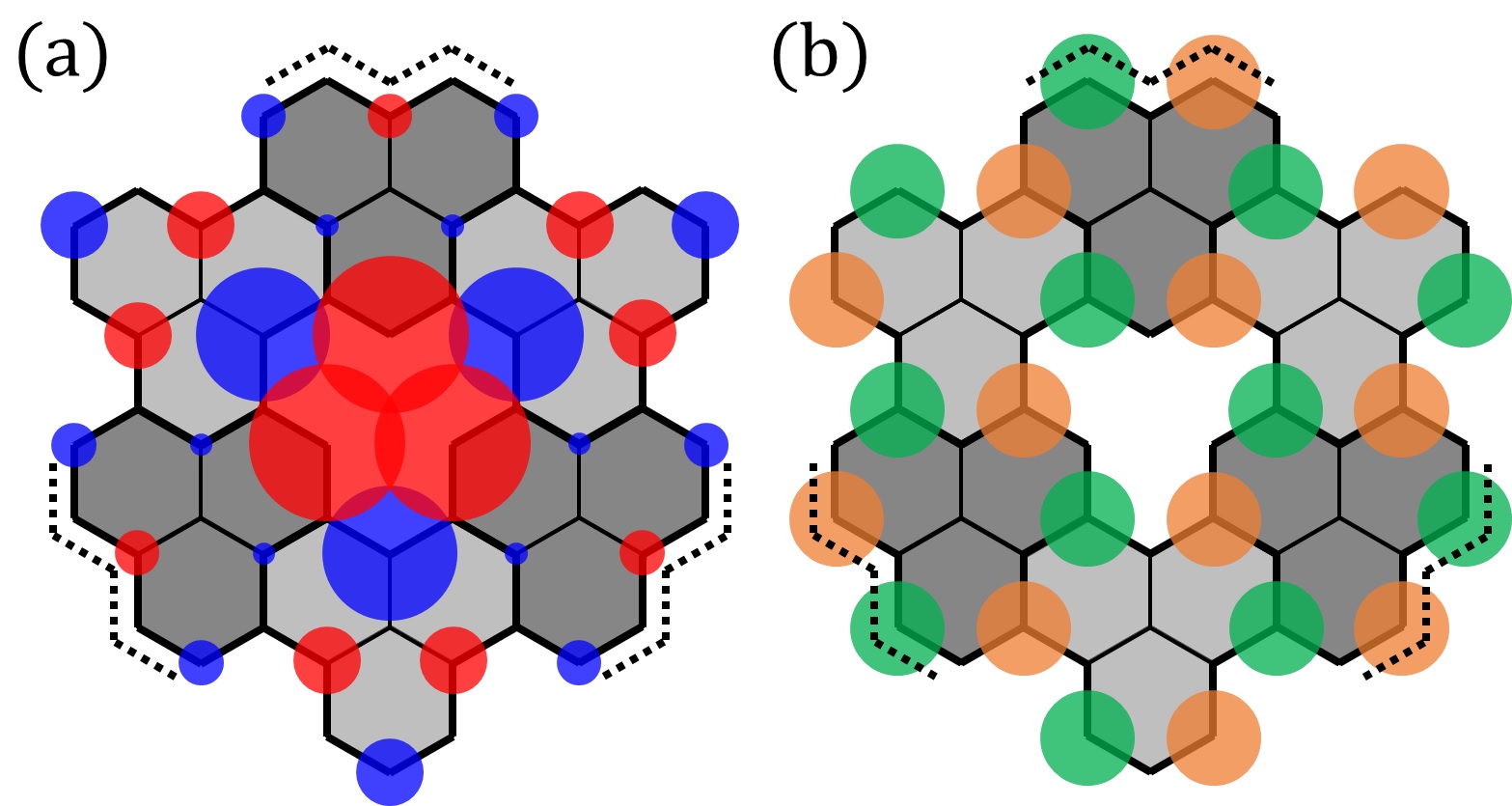}
\caption{\label{PRB_Fig1} 
Vacancy-centered hexagonal armchair nanographene (VANG) is a typical example of a phenalenyl-tessellation molecule (PTM). 
Each dodecagonal carbon skeleton unit colored in light or dark gray is a phenalenyl unit (PU). 
Three double zig-zag corners (DZCs) are indicated by dotted lines. 
(a) A vacancy-originated localized zero mode (red and blue) on the B-sites and (b) a \sqthree -shaped extended Dirac zero mode  (orange and green) on the A-sites are shown. 
The colors of the circles denote the sign of the wave function,  
and their radii show the amplitude of the wave function on each site.
}
\end{figure}

PTMs comprise phenalenyl unit (PU) tiling. 
Figure \ref{PRB_Fig1} shows a typical example (vacancy-centerd hexagonal armchair nanographene, VANG\cite{doi:10.7566/JPSJ.85.084703}).
PTMs have fringes that form a double zig-zag corner (DZC),
where two consecutive zig-zag edges are indicated by dotted lines in Fig. \ref{PRB_Fig1}. 

\begin{figure}[b]
\includegraphics[width=86mm]{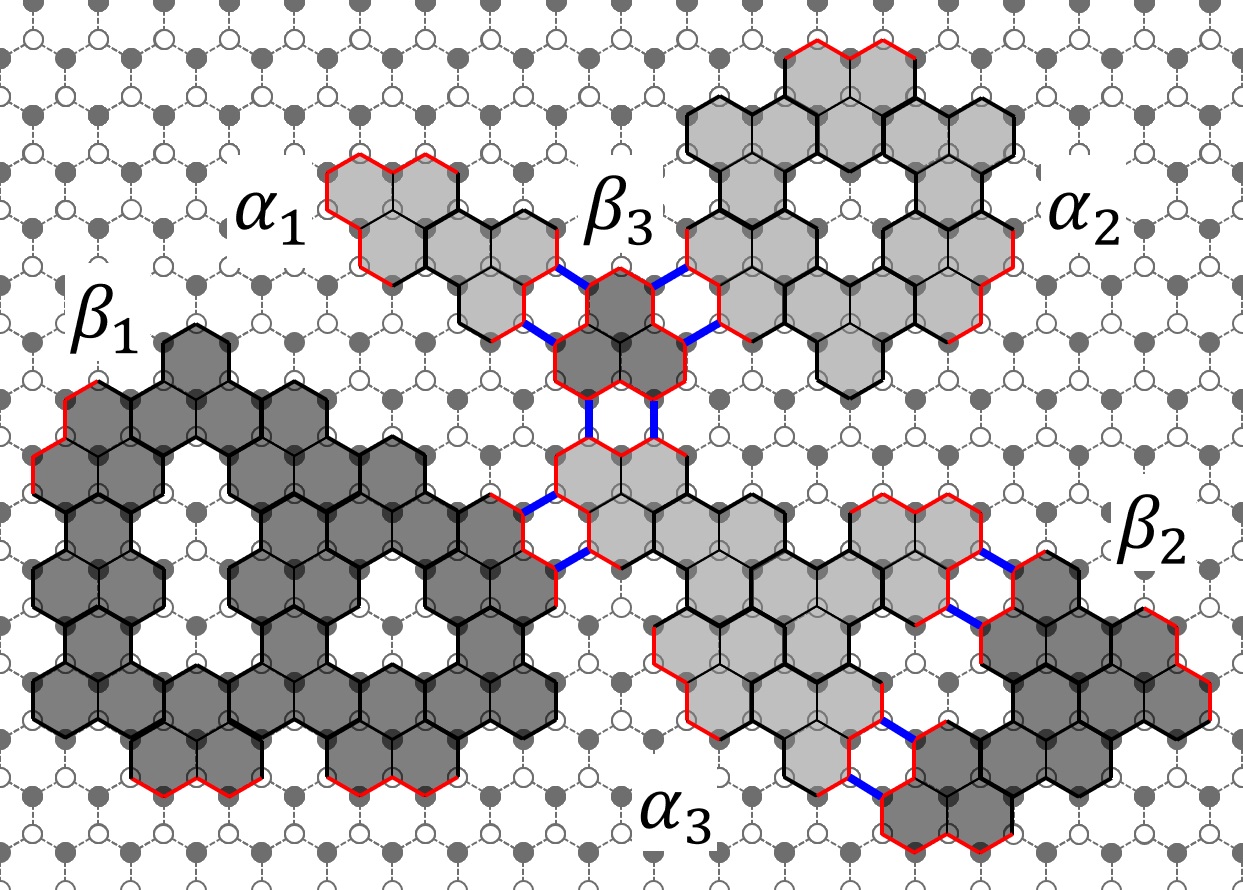}
\caption{\label{PRB_Fig2} 
A polymerized PTM (poly-PTM) system comprising \alpptm s (light gray) and \betptm s (dark gray) on an imaginary honeycomb lattice. 
The two types of PTMs are connected by bonds (blue lines) via their respective DZCs (red lines). 
In the background bipartite honeycomb lattice, the A-sites are colored in white, whereas B-sites are black.
}
\end{figure}

Let us consider the effect that adjusting the atomic positions on an imaginary honeycomb lattice has upon PTMs, as shown in Fig. \ref{PRB_Fig2}. 
PTMs can be categorized into two groups, \alpptm \ and \betptm , 
where the center of each component PU of an \alpptm (a \betptm) is always located at an A-site (a B-site),  
as denoted by white (black) circles. 

In this paper, we consider a polymerized PTM structure of \alp - and \betptm .
In our method for constructing a polymerized PTM, connections ($\sigma$-bonds of carbons) between PTMs are created only at DZCs (shown by blue bonds in Fig. \ref{PRB_Fig2}). 
We call a $\pi$-network system defined in this way a \textit{poly-PTM}. 
We notice that an \alpptm \ and a \betptm \ can be placed in close proximity to each other to create two bonds at their DZCs in a poly-PTM system. 
Two \alpptm s are not directly connected, nor are two \betptm s. 

Let us define the number of $\pi$-$\pi$ connections, $N_\mrm{C}$, between each \alp - and \betptm . 
In Fig. \ref{PRB_Fig2}, $N_\mrm{C}=4$ between the \alpu{3}- and \betu{2}-PTMs, 
whereas $N_\mrm{C}=2$ among the other \alp - and \betptm \ pairs. 

\subsection{Tight-binding model and the zero-sum rule for poly-PTMs}
Below, we consider a single-orbital tight-binding model (TBM) with a constant non-zero nearest neighbor transfer $t \neq 0$ and an on-site energy $\varepsilon = 0$. 
The energy eigen-wavefunction of the TBM is expressed by the amplitude $\psi_{i, l}$ 
at the $i$-th site belonging to the $l$-th sublattice ($l$ = A or B), forming vectors 
$\Psi_\mrm{A} = \trans{(\psi_{1, \mrm{A}}, ..., \psi_{N_\mrm{A}, \mrm{A}})}$ and 
$\Psi_\mrm{B} = \trans{(\psi_{1, \mrm{B}}, ..., \psi_{N_\mrm{B}, \mrm{B}})}$. 
The site index $i$ runs from 1 to $N_\mrm{A}(N_\mrm{B})$ where $N_\mrm{A}(N_\mrm{B})$ is the number of sites in sublattice A(B). 

The Schr\"{o}dinger equation with eigen-energy $E$ is given in matrix form by 
\begin{eqnarray}
\label{eq1}
\left(\begin{array}{cc} 0 & H_\mrm{A \leftarrow B}  \\ H_\mrm{B \leftarrow A} & 0 \\ \end{array} \right)
\left( \begin{array}{cc} \Psi_\mrm{A}\\ \Psi_\mrm{B}\\ \end{array} \right)
=E
\left( \begin{array}{cc} \Psi_\mrm{A}\\ \Psi_\mrm{B}\\ \end{array} \right).
\end{eqnarray}
Here, $H_\mrm{A \leftarrow B}(=H^{\dagger}_\mrm{B \leftarrow A})$ represents a transfer matrix from a B-site to an A-site. 
Equation~(\ref{eq1}) gives the next linear equation when $E=0$: 
\begin{eqnarray}
\label{eq2}
\left\{
\begin{array}{l}
H_\mrm{B \leftarrow A}\Psi_\mrm{A} = 0, \\
H_\mrm{A \leftarrow B}\Psi_\mrm{B} = 0.
\end{array}
\right.
\end{eqnarray}
Equation~(\ref{eq2}) indicates that, when a zero-energy mode $\Psi_\mrm{A(B)}$ with $E=0$ exists on one of A(B)-sites (i.e. the $i_\mrm{A(B)}$-th site), 
the sum of the value of $\Psi_\mrm{B(A)}$ on the B(A)-sites connected to this site must be equal to zero. 
We call this the \textit{zero-sum rule}. 

Conversely, if there exists a wavefunction $\Psi_\mrm{A(B)}$ satisfying the zero-sum rule at each $i_\mrm{B(A)}$-th site in the whole system, 
every element of the resulting vector given by the left side of Eq.~(\ref{eq1}) becomes zero vector.
Then $\Psi_\mrm{A(B)}$ satisfies this equation with the eigenvalue $E=0$. 
Thus, finding a nontrivial wavefunction that satisfies the zero-sum rule in the entire system is equivalent to 
finding the solution of a nontrivial wavefunction $\Psi_\mrm{A(B)}$ with $E=0$ .

Figure \ref{PRB_Fig1} shows examples of the zero modes that satisfy the zero-sum rule on the A(B)-sites. 
One type is a vacancy-originated localized zero mode type, which is found on the B-sites (red and blue), 
and the other is an extended Dirac zero mode, which is found on the A-sites (orange and green).
We can see that the former has larger wavefunction amplitudes at the B-sites around the vacancy than at those on the periphery, 
whereas the latter has perfectly uniform \sqthree -shaped wavefunction in the molecule.


\section{\label{Section3}Results}

\subsection{An example of a poly-PTM}
\begin{figure}[b]
\includegraphics[width=86mm]{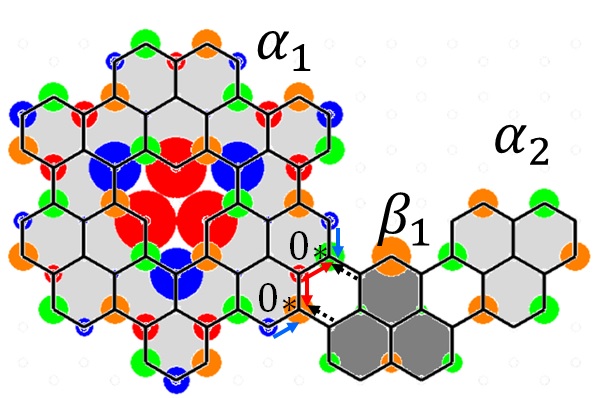}
\caption{\label{PRB_Fig3}
An exemplary poly-PTM comprising \alpu{1}-, \alpu{2}-, and \betu{2}-PTMs. 
We can see that the localized zero mode (LZM) in the \alpu{1}-PTM satisfies the zero-sum rules on the two A-sites marked with stars at the DZC 
by taking the amplitudes of the wavefunction outside the \alpu{1}-PTM to be zero. 
} 
\end{figure}

Figure \ref{PRB_Fig3} is an exemplary poly-PTM comprising two \alpptm s and one \betptm . 
Two types of wavefunctions with $E=0$ can appear in a poly-PTM. 
One is the \textit{localized zero mode (LZM)} type. 
In the example, it is perfectly localized in the \alpu{1}-PTM (red and blue circles on the B-sites). 
Large LZM amplitudes appear mainly around the vacancy of the \alpu{1}-PTM. 
Furthermore, the distribution of the wavefunction amplitudes is not uniform. 
The other type is the \textit{Dirac zero mode (DZM)}; 
in this example, it expands over the \alpu{1}-, \alpu{2}-, and \betu{1}-PTMs (orange and green circles on the A-sites). 
On the A-sites at the \alpu{1}- and \alpu{2}-PTMs, the DZM wavefunction has the special characteristic of uniform \sqthree -shape, 
where the amplitude is zero at the center site of each PU. 
The uniformity of amplitudes is lost on the A-sites in the center, \betu{1}-PTM. 

Such examples obtained by numerical simulations suggest that LZMs appearing in a poly-PTM system
can be characterized as wavefunctions lying on the B(A)-sites of the \alpbetptm s, 
and that each LZM is confined within one of the PTM subgraphs. 
An LZM or a non-uniform part of a DZM can exist on the B(A)-sites of the \alpbetptm s. 
The DZM has a perfectly uniform \sqthree -shaped wavefunction on the counter \betalpptm s. 
To prove these statements, we start by restating rules of the appearance of the zero modes 
on a single isolated PTM and then declare the rules for a general poly-PTM. 


\subsection{PTM Rules}
\begin{theoremP}
\label{theoremP1}
When there are $n$ vacancies in a single isolated \alpbetptm , 
there are $n$ LZMs, which correspond to the vacancy-induced zero modes in the \alpbetptm \ appearing on the B(A)-sites. 
Thus, when there is no vacancy, there is no LZM.
\end{theoremP} 

\begin{theoremP}
\label{theoremP2}
For any single isolated PTM, there is one and only one DZM with a uniform \sqthree -shaped distribution of wavefunction amplitudes. 
The non-zero amplitudes of the DZM are only on the A(B)-sites of the isolated \alpbetptm . 
For each PU, the amplitude of this DZM solution is only zero at the center site; otherwise, it is non-zero.
\end{theoremP} 

\begin{proof}[Proof of PTM Rules \ref{theoremP1} and \ref{theoremP2}]
\label{proofP1P2}
The proof of the PTM Rules can be immediately derived from the proof given in Ref. \cite{doi:10.7566/JPSJ.88.124707} 
and our definition of the A(B)-sites of the \alpbetptm .derived.
\end{proof}

Now, let us consider a poly-PTM system for which there are three rules.


\subsection{Poly-PTM Rules}

\begin{theoremPP}
\label{theoremPP1}
In a poly-PTM, when there are $n$ vacancies in an \alpbetptm ,  
there are $n$ LZMs, which correspond to the vacancy-induced zero modes in the single isolated \alpbetptm appearing on the B(A)-sites.
\end{theoremPP} 

\begin{theoremPP}
\label{theoremPP2}
In a poly-PTM, if there is a wavefunction with zero-energy and 
if and only if non-zero amplitudes of the function are found 
on the A(B)-sites in a \alpbetptm , 
the wavefunction has the \sqthree -shaped distribution within the PTM, 
as described by the PTM Rule 2. 
Thus, the value of the wavefunction 
is represented by a single parameter $\varphi$ as a complex-valued factor in each \alpbetptm .
\end{theoremPP} 

\begin{figure}[t]
\includegraphics[width=86mm]{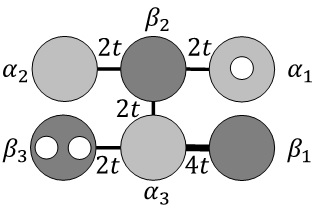}
\caption{\label{PRB_Fig4}
A bipartite super-graph of the poly-PTM system in Fig.~\ref{PRB_Fig2}. 
Each \alpbetptm \ is expressed in light (dark) gray, and the links of the super-graph are black lines. 
Vacancies are indicated by white circles. 
} 
\end{figure}

Here, we introduce a schematic representation of a renormalized graph of the poly-PTM system: a weighted \textit{super-graph}. 
According to our definition of a poly-PTM, the graph is bipartite. 
Figure \ref{PRB_Fig4} shows an exemplary super-graph for the poly-PTM system in Fig. \ref{PRB_Fig2}. 
Each PTM is shown as a node, at which there may be vacancies.  
Using this bipartite super-graph of nodes, it is possible to understand 
how the zero modes appear in a poly-PTM system. 
A generalized zero-sum rule is then naturally introduced, which we call the \textit{super-zero-sum rule}. 
The third rule of poly-PTM governs the DZM's appearance. 

\begin{theoremPP}
\label{theoremPP3}
When $M$ \alp -PTMs and $L$ \bet -PTMs comprise a poly-PTM, 
the connections among them defines a bipartite super-graph of \alp - and \bet -PTMs. 
At each node of the super-graph of \alp - and \bet -PTMs, 
an amplitude $\varphi_{\alpha_i}$ ($\varphi_{\beta_j}$) is assigned, 
where $i=1,\cdots, M$ and $j=1, \cdots, L$. 
Considering the super-zero-sum rule of $\varphi_l$ ($l\in \alpha_i$, or $l\in \beta_j$) on the super-graph, 
if there exists a super-zero mode that satisfies the super-zero-sum rule, then there is a corresponding DZM in the poly-PTM. 
The uniform \sqthree -shaped form of the DZM appears on the A(B)-sites of each \alpbetptm \ subgraph.
\end{theoremPP} 

Proofs of these Poly-PTM Rules are given in the Appendix. 

In summary, the LZMs localized in a PTM appear according to the number of vacancies in each PTM subgraph. 
The DZMs expanding over multiple PTMs appear according to the super-zero-sum rule in the bipartite super-graph of a poly-PTM system made of \alp - and \bet -PTMs.

\section{\label{Section4}Application}

\subsection{Ground states}
\begin{figure}[t]
\includegraphics[width=86mm]{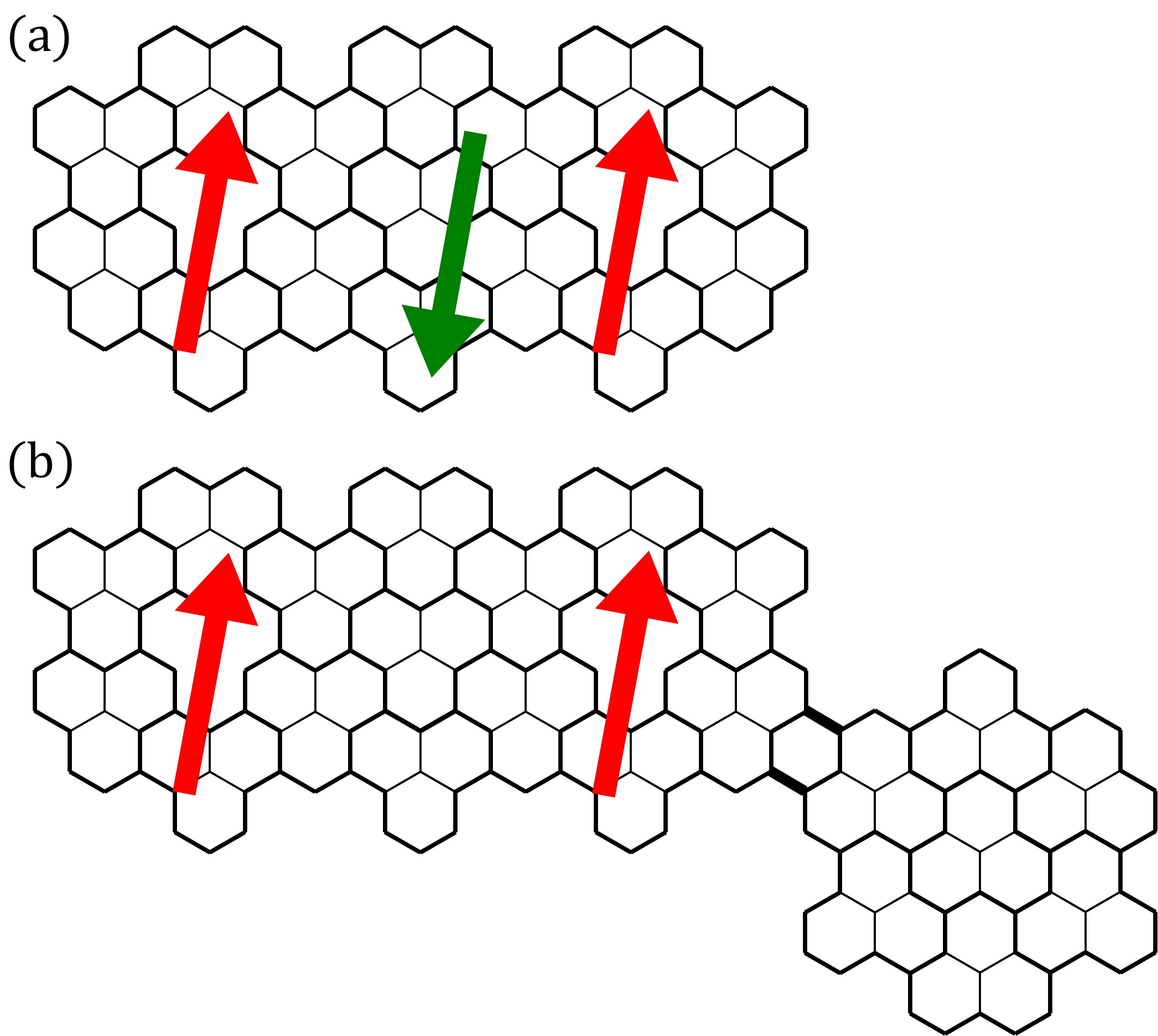}
\caption{\label{PRB_Fig5}
(a) An \alpptm \ with two vacancies. 
Two parallel spins and one anti-parallel spin occur in the two LZMs and one DZM. 
The total ground state becomes a doublet with $S=1/2$. 
(b) A \betptm \ is connected to the \alpptm . 
The DZM disappears, and the ground state becomes triplet with $S=1$. 
} 
\end{figure}

The super-zero-sum rule governing the appearance of DZMs becomes critical when discussing the spin ground state. 
In the following discussion, we follow well-known Ovchinnikov's rule\cite{Ovchinnikov1978} for alternating hydrocarbon systems
and Lieb's theorem\cite{PhysRevLett.62.1201} for repulsive on-site Coulomb interaction within the $\pi$-orbitals.
Typical examples are shown in Fig. ~\ref{PRB_Fig5}. 
Figure ~\ref{PRB_Fig5} (a) presents an \alpptm \ with two vacancies. 
Two LZMs originate from the vacancies, and there is also one DZM from the super-zero-sum rule. 
According to this example, when we focus on the subspace of $E=0$, 
two parallel spins and one anti-parallel spin appear in the two LZMs and the one DZM. 
In the ground state, one of the parallel spins comprises a singlet with the anti-parallel one, 
making the total ground state a doublet for which the total magnetic moment becomes $S=1/2$. 
The same result is mentioned in our previous paper\cite{doi:10.7566/JPSJ.88.124707}. 
In Fig. ~\ref{PRB_Fig5} (b), a \betptm \ is connected to the example mentioned above. 
No DZM occurs according to the super-zero-sum rule; 
therefore, there appear two parallel spins in the LZMs on the B-sites and in the ground state.
These make up a spin triplet with the total magnetic moment of $S=1$. 
These examples show the importance of controlling the appearance of 
both LZMs and DZMs using the super-zero-sum rule to determining the spin ground state 
and the low-energy effective spin Hamiltonian. 
Thus, our rule enables magnetic molecules and magnetic nanocarbon systems to be designed rather freely, 
having advantages compared with other attempts\cite{doi:10.1021/acs.nanolett.9b01773, https://doi.org/10.1002/chem.202003713}. 
This may allow the material realization of quantum-computation-device structures. 

\subsection{Discussion from the viewpoint of the Shima-Aoki theorem}
\begin{figure}[b]
\includegraphics[width=86mm]{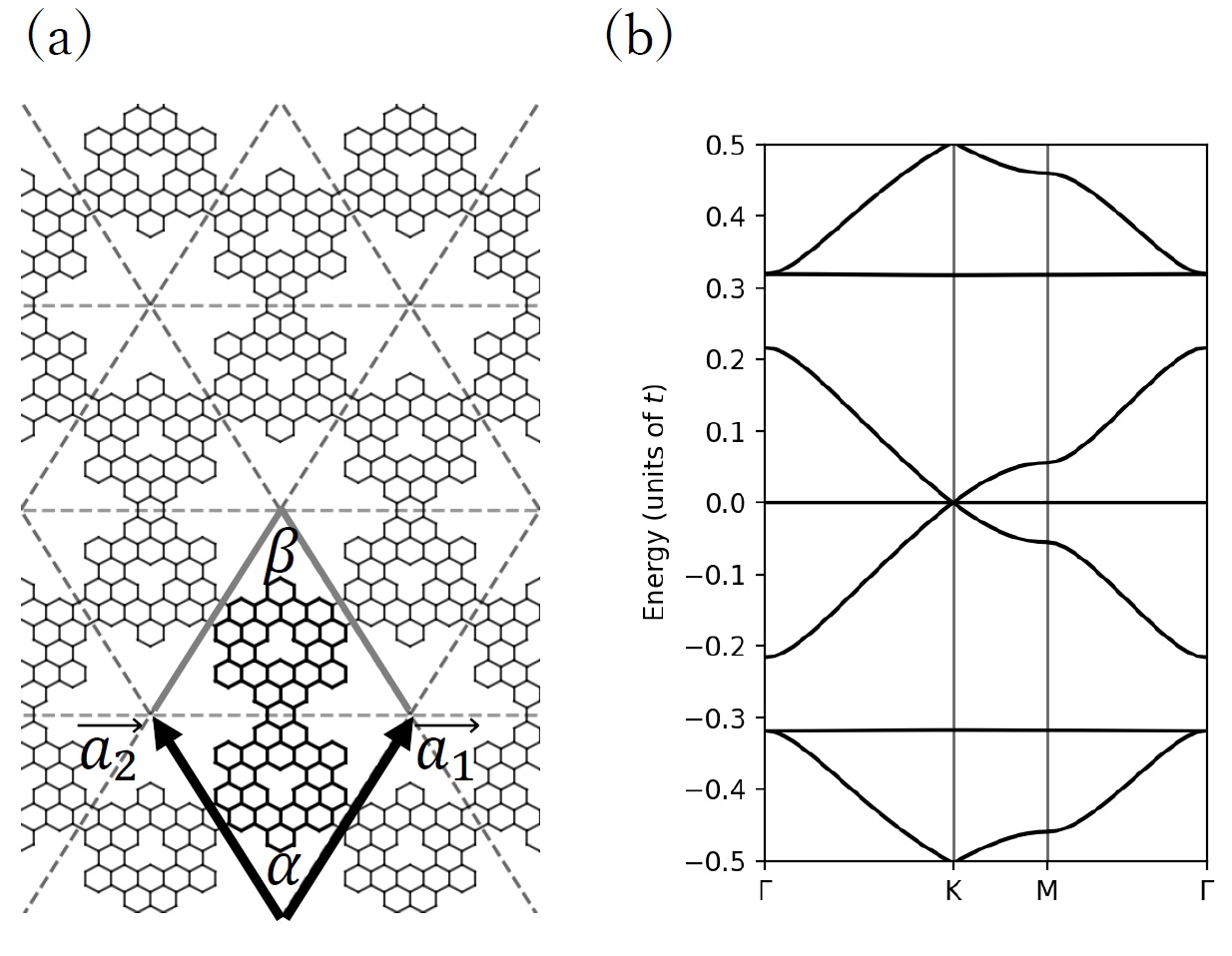}
\caption{\label{PRB_Fig6}
(a) A super-bipartite honeycomb lattice of VANG. Each \alpbetptm is given by a VANG molecule. 
(b) The band structure of the VANG-VANG hexagonal lattice. 
Two flat LZMs degenerate at $E=0$, and the linear DZMs appear according to the super-zero-sum rule rule at the K-point.
} 
\end{figure}

Let us consider the example of a periodic poly-PTM system comprising VANG\cite{doi:10.7566/JPSJ.85.084703,doi:10.7566/JPSJ.86.034802} as \alp - and \betptm .
When VANG forms a two-dimensional hexagonal lattice (Fig. \ref{PRB_Fig6}(a)), 
we have two flat bands at $E=0$ in the band structure (Fig. \ref{PRB_Fig6}(b)), which correspond to the vacancy-induced LZMs. 
There appear two additional zero modes at the K-point in the Brillouin zone derived from the \sqthree -shaped Dirac modes that correspond to the DZMs. 
These have linear dispersions forming Dirac cones 
around the K-point that reflect the graphene-like honeycomb structure of the super-graph. 

In fact, this gapless Dirac branch can be considered an exceptional case of the Shima–Aoki theorem\cite{PhysRevLett.71.4389}. 
When we consider each VANG to be a super-atom in the Shima–Aoki theorem,  
because of there being a vacancy site at each VANG, 
the VANG–VANG super-bipartite honeycomb-lattice structure is classified as type $\mrm{A_0}$\cite{PhysRevLett.71.4389}. 
Although a normal $\mrm{A_0}$ structure should have a semiconducting band gap 
and $m(\ge0)$ flat bands according to the theorem, 
the system may show the gapless Dirac cone,
since an accidental degeneracy is caused by the LZM of the VANG. 
Accidental degeneracy occurs within the zone interior. 
In this system, two zero-energy representations of the $E$ symmetry appear at the K-point. 
These facts by LZMs of VANG forming a zero-energy flat band are really designed by our new Poly-PTM Rules. 
As a result of the super-honeycomb graph of VANG, the gapless Dirac mode remains. 

Of course, we can consider other PTM suructures with $C_3$ symmetry with a vacancy at the center besides a VANG super-atom.
Therefore, there is a scheme for systematically constructing poly-PTM honeycomb structures that constitute exceptional cases of the Shima-Aoki theorem.



\subsection{Construction of a poly-PTM as real-space renormalization group operation}

\begin{figure}[t]
\includegraphics[width=86mm]{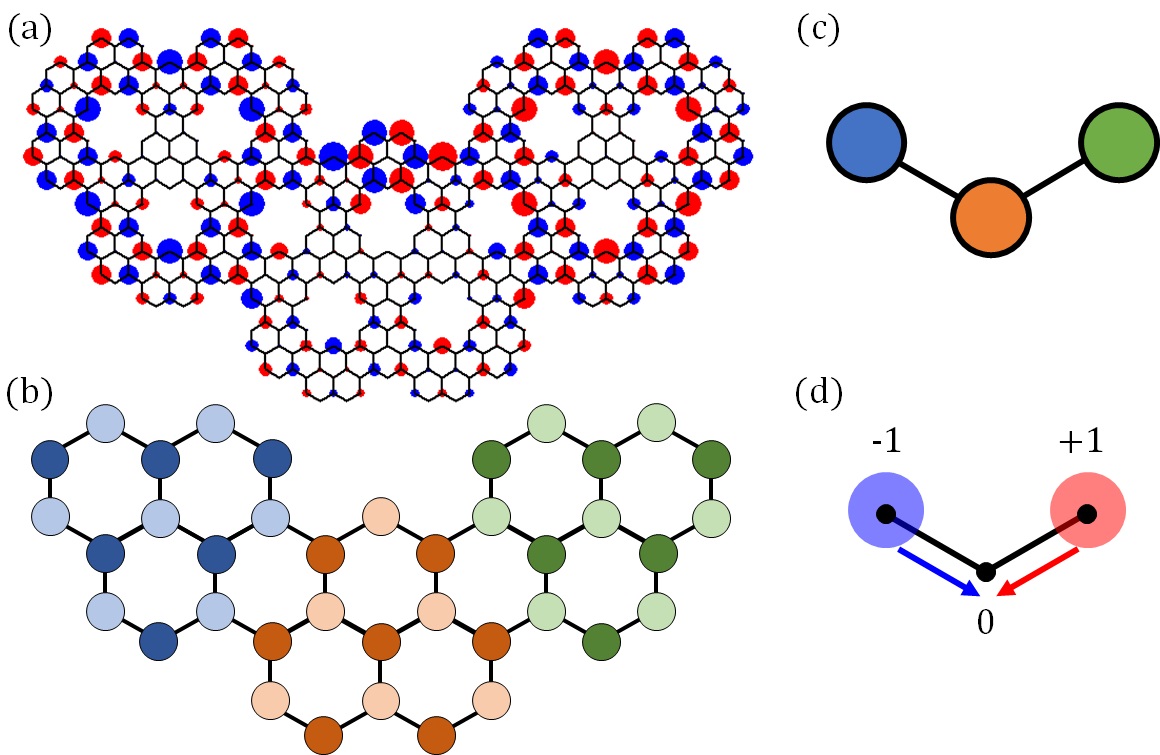}
\caption{\label{PRB_Fig7}
(a) A super-poly-PTM structure; (b)the super-graph; and (c) the corresponding bipartite super-super-graph. 
(d) The zero mode that appears across multiple super-PTMs shown in (a) is governed by a super-super-zero-sum rule. 
} 
\end{figure}

Here, let us mention that the Hamiltonian mapping described in this paper is 
a real-space renormalization-group operation\cite{PhysicsPhysiqueFizika.2.263,RevModPhys.39.395}.
Such an operation can be applied recursively. 
As shown in Fig. \ref{PRB_Fig7}(a), we can construct a \textit{super-poly-PTM} structure in the scope of a poly-PTM with a super-graph (Fig. \ref{PRB_Fig7}(b)), 
and we can consider a corresponding \textit{super-super-graph} (Fig. \ref{PRB_Fig7}(c)). 
A \textit{super-super-zero-sum rule} on a super-super-graph dominates the zero mode that appears across multiple super-PTMs (Fig. \ref{PRB_Fig7}(d)). 
The conclusions of these discussions deepen our understanding of the relationship between the nanostructures and zero modes of graphene. 

\section{\label{Section5}Conclsions}

We found that each PTM with $n$ vacancies will also have  $n$ LZMs 
and that there are also DZMs derived from \sqthree -shaped Dirac modes 
according to the super-zero-sum rule in the bipartite super-graphs composed of multiple PTMs. 
The LZMs are confined for each PTM. 
Based on these rules, it is possible to design a poly-PTM system with multiple LZMs and DZMs in one- and two-dimensional spaces. 
For the poly-PTM structure, previously unknown features, 
such as exceptions to the Shima-Aoki theorem and the possible spatial-renormalization group operations, were revealed. 
This helps lay the groundwork for the design of quantum devices using a two-dimensional honeycomb lattice of graphene. 
From the viewpoint of the spatial-renormalization group for a poly-PTM system, 
it may be interesting to extend this concept to molecules or crystals made of PTMs in three-dimensional space, 
potentially with twisted structures that cannot be realized with a TBM of single carbon atoms. 

\begin{acknowledgments}
The authors thank S. Teranishi and Y. Wicaksono for our illuminating discussions and valuable comments. 
This work is partly supported by KAKENHI No. K034560. 
\end{acknowledgments}


\appendix*
\section{Proofs of Poly-PTM Rules}
\begin{proof}[Proof of Poly-PTM Rule \ref{theoremPP1}]
By letting the amplitudes of the wavefunction be zero outside of the PTM, 
the zero-sum rules on the A(B)-sites at the DZCs, as well as those inside the PTM,  
are satisfied by the $n$ LZMs in the PTM that correspond with those in the isolated PTM. 
An amplitude of zero naturally satisfies the zero-sum rules required on the A(B)-sites outside the PTM, 
such that the $n$ LZMs in the PTM satisfy all zero-sum rules and appear in the poly-PTM system. 
\end{proof}

\begin{proof}[Proof of Poly-PTM Rule \ref{theoremPP2}]
Since each \alpbetptm \ is connected to the outside only at the DZCs, 
the zero-sum rule on each B(A)-site in the \alpbetptm \ does not include the amplitude of the wavefunction on the A(B)-site outside the PTM. 

Suppose that a wavefunction with zero-energy is found in the poly-PTM. 
When its non-zero amplitudes are found on A(B)-sites in one of the \alpbetptm , 
the zero-sum rules on the B(A)-sites in the PTM are satisfied by the wavefunction. 
The zero-sum rule equations are given by a partial set of those in the whole poly-PTM. 
Note that this partial set is identical to the full set of the zero-sum rule equations 
in an isolated \alpbetptm . 
This means that the partial set, a necessary condition for the zero mode of the poly-PTM, 
becomes a sufficient condition for the zero-sum rules of the single isolated PTM, 
but not for the wavefunction to be the zero mode of the poly-PTM. 

Because of PTM Rule 2, 
there can be only \sqthree -shaped Dirac mode on the A(B)-sites in the isolated \alpbetptm \ subgraph, 
which is the unique zero mode appearing on the A(B)-sites of \alpbetptm . 
The shape of the mode is also unique. 
Therefore, the zero mode wavefunction of the poly-PTM has a \sqthree -shaped distribution, 
or zero in each \alpbetptm .
From the uniformity of the \sqthree -shaped distribution, it is clear that the value of the wavefunction 
on each A(B)-site in each \alpbetptm \ can be parametrized by a local phase factor, $\varphi$.
\end{proof}


\begin{figure}[b]
\includegraphics[width=86mm]{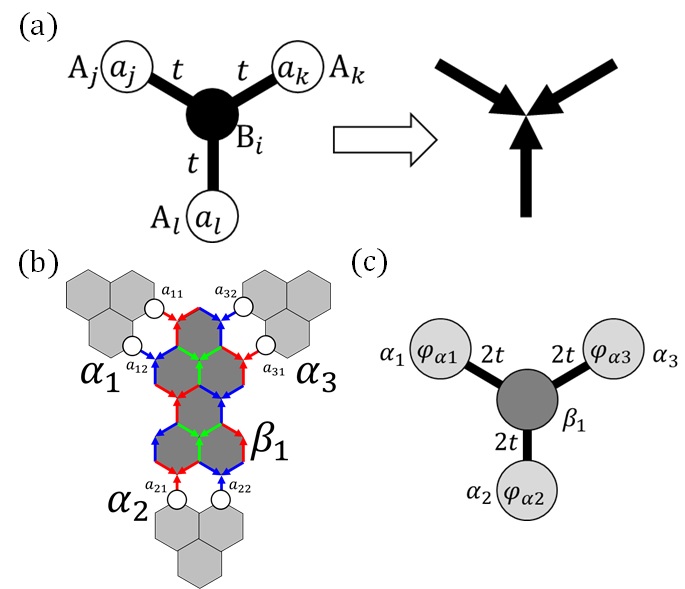}
\caption{\label{PRB_Apdx_Fig1}
(a) Left: zero-sum rule on the $i$-th B-site. Right: The rule is expressed by arrows forming a local arrow diagram. 
(b) Zero-sum rules for the entire \betptm \ are shown by green, red and blue arrows.
(c) Defined super-graph. Each connection is assigned a weight of $2t(=N_\mrm{C}t)$.
} 
\end{figure}

Before moving on to the proof of Poly-PTM Rule \ref{theoremPP3}, 
let us consider an example. 
First, we define an arrow diagram. 
Let us focus upon the wavefunctions on the A-sites. 
The zero-sum rule 
\begin{equation}
\label{eq3}
a_j+a_k+a_l=0
\end{equation}
on each B-site for a wavefunction on the A-site is drawn as shown on the left-hand side of Fig. \ref{PRB_Apdx_Fig1}(a). 
Here, $a_j$, $a_k$, and $a_l$ represent the values of the wavefunction 
on the $j$-, $k$-, and $l$-th A-sites around the $i$-th B-site.
The right-hand side of Fig. \ref{PRB_Apdx_Fig1}(a) shows an arrow diagram: three arrows are pointing to the $i$-th B-site. 
Using this structure as a local diagram, we can write the whole zero-sum rule in a global diagram. 

A poly-PTM is shown in Fig. \ref{PRB_Apdx_Fig1}(b). 
In this system, we have two zero modes on the A-site of the poly-PTM, 
which are finally found to be DZMs satisfying the zero-sum rules at the B-sites. 
We focus upon the \betptm \ at the center of the network; 
then, the zero-sum rules of the entire \betptm \ can be drawn by the arrow diagram. 
The zero-sum rule at the center of each PU is colored green. 
Then, we see that the zero-sum rules at the periphery of each PU can be alternately colored red and blue. 
Consequently, the edges of the \betptm \ are always colored red and blue as well. 
This is a common property of general PTM. 

In this example, there are 11 A-sites and 12 B-sites in the \betu{1}-PTM. 
When we consider the zero-sum rules at every B-sites in the \betu{1}-PTM, 
we obtain 12 linearly independent equations, which are linear equations of the zero mode wavefunction. 
The values of these modes are indicated as $a_{\beta_1,l}$ with $l=1, \cdots, N_{\beta,\mrm{A}}=11$ for the 11 A-sites. 
In the zero-sum rules, there appear six more mode values. 
In the adjacent \alpu{1}-, \alpu{2}- and \alpu{3}-PTMs, there are $2\times 3=6$ A-sites, 
for which the values of the mode contribute to the zero-sum rules. 
Therefore, the six additional values indicated by $a_{\alpha_i,l}\equiv a_{il}$ appear with $i=1,\cdots,M=3$. 
Relevant A-sites in each \alpptm \ are those at a DZC. 
Thus, we renumber the A-sites in \alpptm , and $l=1,2$ is considered as shown in Fig. \ref{PRB_Apdx_Fig1}(b). 
The zero-sum rules contain 17 variables (i.e. $a_{\beta_1,l}$ and $a_{il}$ for $17(=11+6)$) A-sites. 

We will show that, when the values of the zero mode at 
the exterior A-sites in the adjacent \alpptm s (i.e. \alpu{1}-, \alpu{2}-, and \alpu{3}-PTMs) 
are determined as boundary conditions for the interior, 
$a_{\beta_1,l}$ can be uniquely determined to obtain the full solution. 
The wavefunction of the \alpptms \ can be determined using the {\it super-zero-sum rule} introduced below. 

Here, under Poly-PTM Rule \ref{theoremPP2}, 
we can concisely express the relative value of the \sqthree -shaped Dirac mode in each \alpu{i} PTM. 
Considering the possible DZM on A(B)-sites, 
let us introcduce a number $\varphi_{\alpha(\beta)i}$ to each \alpbetptm , where $\varphi_{\alpha(\beta)i}$ is complex-valued and can be zero. 
This value represents a factor of the \sqthree -shaped Dirac mode on the A(B)-sites in each \alpbetptm . 
Then, we can also derive the following relationships from the \sqthree \ nature of the Dirac mode on the adjacent \alpu{1}-, \alpu{2}- and \alpu{3}-PTMs:
\begin{eqnarray}
\label{eq4}
\left\{
\begin{array}{l}
a_{11} = -a_{12} \coloneqq \varphi_{\alpha_1}, \\
a_{21} = -a_{22} \coloneqq \varphi_{\alpha_2}, \\
a_{31} = -a_{32} \coloneqq \varphi_{\alpha_3}.
\end{array}
\right.
\end{eqnarray}
Since these three conditions are determined by the zero-sum rules in the \alpu{i}-PTMs, 
the constraints are independent of the 12 above-mentioned equations. 
Therefore, the number of degrees of freedom (DOF) 
remaining for the wavefunction with respect to $a_{\beta_1,l}$ and $a_{il}$ is $17-12-3=2$.

The problem of determining the number of DOF can be translated into a closed set of equations among the amplitudes 
of the wavefunctions of the adjacent \alpu{1}-, \alpu{2}- and \alpu{3}-PTMs, 
as expressed by $\varphi_{\alpha_i}$. 
By adding each side of the corresponding zero-sum rule [Eq.~(\ref{eq3})] shown in a red diagram
and subtracting each zero-sum rule equation shown in the blue diagram, we obtain: 
\begin{equation}
\label{eq5}
a_{11}+a_{21}+a_{31}-(a_{12}+a_{22}+a_{32})=0.
\end{equation}
This simplification is always possible because the $a_{\beta_1,l}$ at the A-sites on the periphery of 
all PUs in the \betu{1}-PTM appears twice in zero-sum rule equations at the B-sites on the periphery: 
once in the equation for the red-colored diagram and again for the blue-colored diagram. 
Therefore, adding the equations with alternating positive and negative factors 
along the circumference of PUs into the \betu{1}-PTM  
cancels-out $a_{\beta_1,l}$ in the final expression. 
Along the way, $a_{i1}$ and $a_{i2}$ remain for concluding Eq.~(\ref{eq5}). 

From Eqs. (\ref{eq4}) and (\ref{eq5}), we can derive the following relationship: 
\begin{equation}
\label{eq6}
2\varphi_{\alpha_1}+2\varphi_{\alpha_2}+2\varphi_{\alpha_3}=0.
\end{equation}
Note that the coefficient for each $\varphi$ in Eq.~(\ref{eq6}) corresponds to $N_\mrm{C}=2$. 

For the present example, there are two linearly independent solutions satisfying Eq.~(\ref{eq6}). 
Indeed, we know that the three-dimensional vectors following one constraint making the vector 
be normal to one direction are given in a two-dimensional vector space. 
Consequently, there are two DZMs. 
Here, we need to certify the existence of a unique solution 
for each of determined set of $\varphi_{\alpha_i}$ ($i=1,2$). 
This is really the case, because, except for Eq.~(\ref{eq6}), 
11 simultaneous linear equations forming a matrix equation $X\bm{a} = \bm{b}$ 
remains for the determination of $\bm{a} = \trans{(a_{\beta_1,1},\cdots,a_{\beta_1,11})}$, 
where a non-trivial vector $\bm{b}$ is given by $\varphi_{\alpha_i}$. 
The matrix $X$ is regular because the transformation from the original 12 zero-sum rule equations is linear. 

Equation ~(\ref{eq6}) expresses the relationship among $\varphi_{\alpha_1}$, $\varphi_{\alpha_2}$ and  $\varphi_{\alpha_3}$ with DOF of two for the values of the  wavefunctions on the \alpu{1}-, \alpu{2}-, and \alpu{3}-PTMs connected to the \betu{1}-PTM. 
This can be regarded as a \textit{super-zero-sum rule} on the \textit{super-graph} defined by \alpu{1}-, \alpu{2}-, \alpu{3}-, and \betu{1}-PTMs and the connections among them, as shown in Fig. \ref{PRB_Apdx_Fig1}(c).
Two linearly independent solutions of the super-zero-sum rule expressed in Eq. ~(\ref{eq6}) are the \textit{super-zero modes} that appear in the system shown in Fig. \ref{PRB_Apdx_Fig1}(c).  
Using this super-zero-sum rule, we can make a proof of Poly-PTM Rule \ref{theoremPP3}. 


\begin{proof}[Proof of Poly-PTM Rule \ref{theoremPP3}]
Let us consider a \betptm \ with $n$ vacancies. 
Suppose that this PTM is connected to $M(> 1)$ adjacent \alpptm s.
Let $N_\mrm{C}$ among them be $N_{\mrm{C}_1}$, $N_{\mrm{C}_2}$, ..., $N_{\mrm{C}_M}$. 
There are $N_{\beta,\mrm{A}}$ A-sites in the \betptm ; 
in the adjacent \alpptm s, there are $\sum_{i=1}^{M}N_{\mrm{C}_i}$ A-sites that connect to the DZCs of the \betptm .
When we focus upon the wavefunctions on the A-sites, 
there are $N_{\beta, \mrm{B}}$ linearly independent zero-sum rule equations on the B-sites in the \betptm ; 
by the Poly-PTM Rule \ref{theoremPP2}, 
there are $\sum_{i=1}^{M} (N_{\mrm{C}_i}-1)$ linearly independent equations from the \sqthree \ nature of the Dirac mode on the adjacent \alpptm s. 
In total, the wavefunction on the A-sites has 
$N_{\beta,\mrm{A}}+\sum_{i=1}^{M}N_{\mrm{C}_i}-N_{\beta,\mrm{B}}-\sum_{i=1}^{M} (N_{\mrm{C}_i}-1)=n+M-1$ DOFs. 
Here, we used the relationship $N_{\beta,\mrm{A}}-N_{\beta,\mrm{B}}=n-1$ between the number of the A-sites, 
B-sites, and vacancies in the \betptm \cite{doi:10.7566/JPSJ.88.124707}. 
Since  $n$ DOFs originate from the vacancies corresponds to the LZMs, 
the remaining $M-1$ DOFs correspond to DZMs. 
Then, by considering colored arrow diagram as shown in Fig. \ref{PRB_Apdx_Fig1}(b) and 
adding each side of the zero-sum rule equations in red and subtracting those blue,  
the relationship among the \sqthree -shaped Dirac wavefunctions on the adjacent \alpptm s becomes as follows: 
\begin{equation}
\label{eq7}
t\sum_{i=1}^{M} N_{\mrm{C}_i}\varphi_{\alpha_i}=0.
\end{equation}
When we focus upon the wavefunctions on the B-sites in an \alpptm \ which has $L$ adjacent \betptm s, 
based the same discussion, the relationship among the \sqthree -shaped Dirac modes also becomes 
\begin{equation}
\label{eq8}
t\sum_{j=1}^{L} N_{\mrm{C}_j}\varphi_{\beta_j}=0.
\end{equation}

When we consider a poly-PTM comprising $N_\alpha$ \alpptm s and $N_\beta$ \betptm s, 
we can define a super-graph corresponding to the poly-PTM structure 
by assigning a weight $tN_{\mrm{C}_{k-l}}$ to each link between the $k$-th \alpptm \ and the $l$-th \betptm , 
where $N_{\mrm{C}_{k-l}}$ is the number of the connections. 

By considering Eqs. (\ref{eq7}) and (\ref{eq8}) on all \bet - and \alp-PTMs in a poly-PTM, 
respectively, we obtain the following equations: 
\begin{eqnarray}
\label{eq9}
\left\{
\begin{array}{l}
H_{\beta \leftarrow \alpha}\Phi_{\alpha} = 0, \\
H_{\alpha \leftarrow \beta}\Phi_{\beta} = 0.
\end{array}
\right.
\end{eqnarray}
Here, $\Phi_{\alpha}=\trans{(\varphi_{\alpha_{1}}, ..., \varphi_{\alpha_{N_{\alpha}}})}$ and $\Phi_{\beta}=\trans{(\varphi_{\beta_{1}}, ..., \varphi_{\beta_{N_{\beta}}})}$.
$H_{\beta \leftarrow \alpha}(=H^{\dagger}_{\alpha \leftarrow \beta})$ can be seen as an effective transfer matrix from \alpbetptm \ to a \betalpptm , 
which is defined as $(H_{\beta \leftarrow \alpha})_{lk}=tN_{\mrm{C}_{k-l}}$. 
Equation (\ref{eq9}) expresses relationships among the values of the \sqthree -shaped wavefunctions on the \alp - and \betptm s in the poly-PTM. 
This is the super-zero-sum rule on the super-graph, and the solution is the super-zero mode.

According to the zero-sum rule, the determination equations for the wavefunctions at the A(B)-sites in the \betalpptm \ subgraph are mutually independent. 
For example, for the A-sites in the \betptm \ subgraph, 
we have $N_{\beta,\mrm{B}} = N_{\beta,\mrm{A}} - n + 1 \leq N_{\beta,\mrm{A}}+1$. 
There appears to be one super-zero-sum rule equation per \betptm \ according to a linear transformation. 
The other $N_{\beta,\mrm{A}} - n$ equations determine the wavefunction on these sites 
since a set of simultaneous linear equations can be solved if the number of equations 
is not larger than the number of variables. 

So, if Eq. (\ref{eq9}) has a non-trivial solution as a super-zero mode, 
there exists a corresponding DZM in the poly-PTM, 
which has a uniform \sqthree -shaped Dirac mode at the A(B)-sites in each \alpbetptm \ subgraph.
\end{proof}

The kernel of the original poly-PTM is governed by the super-zero-sum rule. 
In fact, any non-trivial solution of Eq. (\ref{eq9}) as the kernel of the effective problem has 
a corresponding solution on the poly-PTM. 
Note that the converse is not true, since a LZM corresponds to none of the solutions of Eq. (\ref{eq9}).

The super-zero-sum rule in Eq. (\ref{eq9}) certifies existence of chirality in the super-graph. 
Indeed, we arrived at a renormalized equation determining zero modes. 
The super-graph is defined as the renormalized bipartite graph obtained from the original poly-PTM graph. 
The same super-graph can be obtained from microscopically different poly-PTM graphs. 
However, the kernels have the same dimension as far as the Dirac zero modes are considered. 
There is a one-to-one correspondence between a super-zero mode of the super-graph and a Dirac zero mode in the poly-PTM. 
This is possible since we are looking only at a special subset of the zero-energy spectrum. 
A microscopic LZM may be added;  
however, we believe that the present result may open a scope also for the renormalization group. 




\bibliographystyle{apsrev4-2}

\begin{thebibliography}{36}%
\makeatletter
\providecommand \@ifxundefined [1]{%
 \@ifx{#1\undefined}
}%
\providecommand \@ifnum [1]{%
 \ifnum #1\expandafter \@firstoftwo
 \else \expandafter \@secondoftwo
 \fi
}%
\providecommand \@ifx [1]{%
 \ifx #1\expandafter \@firstoftwo
 \else \expandafter \@secondoftwo
 \fi
}%
\providecommand \natexlab [1]{#1}%
\providecommand \enquote  [1]{``#1''}%
\providecommand \bibnamefont  [1]{#1}%
\providecommand \bibfnamefont [1]{#1}%
\providecommand \citenamefont [1]{#1}%
\providecommand \href@noop [0]{\@secondoftwo}%
\providecommand \href [0]{\begingroup \@sanitize@url \@href}%
\providecommand \@href[1]{\@@startlink{#1}\@@href}%
\providecommand \@@href[1]{\endgroup#1\@@endlink}%
\providecommand \@sanitize@url [0]{\catcode `\\12\catcode `\$12\catcode
  `\&12\catcode `\#12\catcode `\^12\catcode `\_12\catcode `\%12\relax}%
\providecommand \@@startlink[1]{}%
\providecommand \@@endlink[0]{}%
\providecommand \url  [0]{\begingroup\@sanitize@url \@url }%
\providecommand \@url [1]{\endgroup\@href {#1}{\urlprefix }}%
\providecommand \urlprefix  [0]{URL }%
\providecommand \Eprint [0]{\href }%
\providecommand \doibase [0]{https://doi.org/}%
\providecommand \selectlanguage [0]{\@gobble}%
\providecommand \bibinfo  [0]{\@secondoftwo}%
\providecommand \bibfield  [0]{\@secondoftwo}%
\providecommand \translation [1]{[#1]}%
\providecommand \BibitemOpen [0]{}%
\providecommand \bibitemStop [0]{}%
\providecommand \bibitemNoStop [0]{.\EOS\space}%
\providecommand \EOS [0]{\spacefactor3000\relax}%
\providecommand \BibitemShut  [1]{\csname bibitem#1\endcsname}%
\let\auto@bib@innerbib\@empty
\bibitem [{\citenamefont {Novoselov}\ \emph {et~al.}(2004)\citenamefont
  {Novoselov}, \citenamefont {Geim}, \citenamefont {Morozov}, \citenamefont
  {Jiang}, \citenamefont {Zhang}, \citenamefont {Dubonos}, \citenamefont
  {Grigorieva},\ and\ \citenamefont {Firsov}}]{Novoselov666}%
  \BibitemOpen
  \bibfield  {author} {\bibinfo {author} {\bibfnamefont {K.~S.}\ \bibnamefont
  {Novoselov}}, \bibinfo {author} {\bibfnamefont {A.~K.}\ \bibnamefont {Geim}},
  \bibinfo {author} {\bibfnamefont {S.~V.}\ \bibnamefont {Morozov}}, \bibinfo
  {author} {\bibfnamefont {D.}~\bibnamefont {Jiang}}, \bibinfo {author}
  {\bibfnamefont {Y.}~\bibnamefont {Zhang}}, \bibinfo {author} {\bibfnamefont
  {S.~V.}\ \bibnamefont {Dubonos}}, \bibinfo {author} {\bibfnamefont {I.~V.}\
  \bibnamefont {Grigorieva}},\ and\ \bibinfo {author} {\bibfnamefont {A.~A.}\
  \bibnamefont {Firsov}},\ }\href {https://doi.org/10.1126/science.1102896}
  {\bibfield  {journal} {\bibinfo  {journal} {Science}\ }\textbf {\bibinfo
  {volume} {306}},\ \bibinfo {pages} {666} (\bibinfo {year}
  {2004})}\BibitemShut {NoStop}%
\bibitem [{\citenamefont {Geim}\ and\ \citenamefont
  {Novoselov}(2007)}]{Geim2007}%
  \BibitemOpen
  \bibfield  {author} {\bibinfo {author} {\bibfnamefont {A.~K.}\ \bibnamefont
  {Geim}}\ and\ \bibinfo {author} {\bibfnamefont {K.~S.}\ \bibnamefont
  {Novoselov}},\ }\href {https://doi.org/10.1038/nmat1849} {\bibfield
  {journal} {\bibinfo  {journal} {Nature Materials}\ }\textbf {\bibinfo
  {volume} {6}},\ \bibinfo {pages} {183} (\bibinfo {year} {2007})}\BibitemShut
  {NoStop}%
\bibitem [{\citenamefont {Novoselov}\ \emph {et~al.}(2005)\citenamefont
  {Novoselov}, \citenamefont {Geim}, \citenamefont {Morozov}, \citenamefont
  {Jiang}, \citenamefont {Katsnelson}, \citenamefont {Grigorieva},
  \citenamefont {Dubonos},\ and\ \citenamefont {Firsov}}]{Novoselov2005}%
  \BibitemOpen
  \bibfield  {author} {\bibinfo {author} {\bibfnamefont {K.~S.}\ \bibnamefont
  {Novoselov}}, \bibinfo {author} {\bibfnamefont {A.~K.}\ \bibnamefont {Geim}},
  \bibinfo {author} {\bibfnamefont {S.~V.}\ \bibnamefont {Morozov}}, \bibinfo
  {author} {\bibfnamefont {D.}~\bibnamefont {Jiang}}, \bibinfo {author}
  {\bibfnamefont {M.~I.}\ \bibnamefont {Katsnelson}}, \bibinfo {author}
  {\bibfnamefont {I.~V.}\ \bibnamefont {Grigorieva}}, \bibinfo {author}
  {\bibfnamefont {S.~V.}\ \bibnamefont {Dubonos}},\ and\ \bibinfo {author}
  {\bibfnamefont {A.~A.}\ \bibnamefont {Firsov}},\ }\href
  {https://doi.org/10.1038/nature04233} {\bibfield  {journal} {\bibinfo
  {journal} {Nature}\ }\textbf {\bibinfo {volume} {438}},\ \bibinfo {pages}
  {197} (\bibinfo {year} {2005})}\BibitemShut {NoStop}%
\bibitem [{\citenamefont {Castro~Neto}\ \emph {et~al.}(2009)\citenamefont
  {Castro~Neto}, \citenamefont {Guinea}, \citenamefont {Peres}, \citenamefont
  {Novoselov},\ and\ \citenamefont {Geim}}]{RevModPhys.81.109}%
  \BibitemOpen
  \bibfield  {author} {\bibinfo {author} {\bibfnamefont {A.~H.}\ \bibnamefont
  {Castro~Neto}}, \bibinfo {author} {\bibfnamefont {F.}~\bibnamefont {Guinea}},
  \bibinfo {author} {\bibfnamefont {N.~M.~R.}\ \bibnamefont {Peres}}, \bibinfo
  {author} {\bibfnamefont {K.~S.}\ \bibnamefont {Novoselov}},\ and\ \bibinfo
  {author} {\bibfnamefont {A.~K.}\ \bibnamefont {Geim}},\ }\href
  {https://doi.org/10.1103/RevModPhys.81.109} {\bibfield  {journal} {\bibinfo
  {journal} {Rev. Mod. Phys.}\ }\textbf {\bibinfo {volume} {81}},\ \bibinfo
  {pages} {109} (\bibinfo {year} {2009})}\BibitemShut {NoStop}%
\bibitem [{\citenamefont {Fujita}\ \emph {et~al.}(1996)\citenamefont {Fujita},
  \citenamefont {Wakabayashi}, \citenamefont {Nakada},\ and\ \citenamefont
  {Kusakabe}}]{doi:10.1143/JPSJ.65.1920}%
  \BibitemOpen
  \bibfield  {author} {\bibinfo {author} {\bibfnamefont {M.}~\bibnamefont
  {Fujita}}, \bibinfo {author} {\bibfnamefont {K.}~\bibnamefont {Wakabayashi}},
  \bibinfo {author} {\bibfnamefont {K.}~\bibnamefont {Nakada}},\ and\ \bibinfo
  {author} {\bibfnamefont {K.}~\bibnamefont {Kusakabe}},\ }\href
  {https://doi.org/10.1143/JPSJ.65.1920} {\bibfield  {journal} {\bibinfo
  {journal} {Journal of the Physical Society of Japan}\ }\textbf {\bibinfo
  {volume} {65}},\ \bibinfo {pages} {1920} (\bibinfo {year}
  {1996})}\BibitemShut {NoStop}%
\bibitem [{\citenamefont {Nakada}\ \emph {et~al.}(1996)\citenamefont {Nakada},
  \citenamefont {Fujita}, \citenamefont {Dresselhaus},\ and\ \citenamefont
  {Dresselhaus}}]{PhysRevB.54.17954}%
  \BibitemOpen
  \bibfield  {author} {\bibinfo {author} {\bibfnamefont {K.}~\bibnamefont
  {Nakada}}, \bibinfo {author} {\bibfnamefont {M.}~\bibnamefont {Fujita}},
  \bibinfo {author} {\bibfnamefont {G.}~\bibnamefont {Dresselhaus}},\ and\
  \bibinfo {author} {\bibfnamefont {M.~S.}\ \bibnamefont {Dresselhaus}},\
  }\href {https://doi.org/10.1103/PhysRevB.54.17954} {\bibfield  {journal}
  {\bibinfo  {journal} {Phys. Rev. B}\ }\textbf {\bibinfo {volume} {54}},\
  \bibinfo {pages} {17954} (\bibinfo {year} {1996})}\BibitemShut {NoStop}%
\bibitem [{\citenamefont {Ryu}\ and\ \citenamefont
  {Hatsugai}(2002)}]{PhysRevLett.89.077002}%
  \BibitemOpen
  \bibfield  {author} {\bibinfo {author} {\bibfnamefont {S.}~\bibnamefont
  {Ryu}}\ and\ \bibinfo {author} {\bibfnamefont {Y.}~\bibnamefont {Hatsugai}},\
  }\href {https://doi.org/10.1103/PhysRevLett.89.077002} {\bibfield  {journal}
  {\bibinfo  {journal} {Phys. Rev. Lett.}\ }\textbf {\bibinfo {volume} {89}},\
  \bibinfo {pages} {077002} (\bibinfo {year} {2002})}\BibitemShut {NoStop}%
\bibitem [{\citenamefont {Klusek}\ \emph {et~al.}(2000)\citenamefont {Klusek},
  \citenamefont {Waqar}, \citenamefont {Denisov}, \citenamefont {Kompaniets},
  \citenamefont {Makarenko}, \citenamefont {Titkov},\ and\ \citenamefont
  {Bhatti}}]{KLUSEK2000508}%
  \BibitemOpen
  \bibfield  {author} {\bibinfo {author} {\bibfnamefont {Z.}~\bibnamefont
  {Klusek}}, \bibinfo {author} {\bibfnamefont {Z.}~\bibnamefont {Waqar}},
  \bibinfo {author} {\bibfnamefont {E.}~\bibnamefont {Denisov}}, \bibinfo
  {author} {\bibfnamefont {T.}~\bibnamefont {Kompaniets}}, \bibinfo {author}
  {\bibfnamefont {I.}~\bibnamefont {Makarenko}}, \bibinfo {author}
  {\bibfnamefont {A.}~\bibnamefont {Titkov}},\ and\ \bibinfo {author}
  {\bibfnamefont {A.}~\bibnamefont {Bhatti}},\ }\href
  {https://doi.org/https://doi.org/10.1016/S0169-4332(00)00374-3} {\bibfield
  {journal} {\bibinfo  {journal} {Applied Surface Science}\ }\textbf {\bibinfo
  {volume} {161}},\ \bibinfo {pages} {508 } (\bibinfo {year}
  {2000})}\BibitemShut {NoStop}%
\bibitem [{\citenamefont {Kobayashi}\ \emph {et~al.}(2005)\citenamefont
  {Kobayashi}, \citenamefont {Fukui}, \citenamefont {Enoki}, \citenamefont
  {Kusakabe},\ and\ \citenamefont {Kaburagi}}]{PhysRevB.71.193406}%
  \BibitemOpen
  \bibfield  {author} {\bibinfo {author} {\bibfnamefont {Y.}~\bibnamefont
  {Kobayashi}}, \bibinfo {author} {\bibfnamefont {K.-i.}\ \bibnamefont
  {Fukui}}, \bibinfo {author} {\bibfnamefont {T.}~\bibnamefont {Enoki}},
  \bibinfo {author} {\bibfnamefont {K.}~\bibnamefont {Kusakabe}},\ and\
  \bibinfo {author} {\bibfnamefont {Y.}~\bibnamefont {Kaburagi}},\ }\href
  {https://doi.org/10.1103/PhysRevB.71.193406} {\bibfield  {journal} {\bibinfo
  {journal} {Phys. Rev. B}\ }\textbf {\bibinfo {volume} {71}},\ \bibinfo
  {pages} {193406} (\bibinfo {year} {2005})}\BibitemShut {NoStop}%
\bibitem [{\citenamefont {Niimi}\ \emph {et~al.}(2005)\citenamefont {Niimi},
  \citenamefont {Matsui}, \citenamefont {Kambara}, \citenamefont {Tagami},
  \citenamefont {Tsukada},\ and\ \citenamefont {Fukuyama}}]{NIIMI200543}%
  \BibitemOpen
  \bibfield  {author} {\bibinfo {author} {\bibfnamefont {Y.}~\bibnamefont
  {Niimi}}, \bibinfo {author} {\bibfnamefont {T.}~\bibnamefont {Matsui}},
  \bibinfo {author} {\bibfnamefont {H.}~\bibnamefont {Kambara}}, \bibinfo
  {author} {\bibfnamefont {K.}~\bibnamefont {Tagami}}, \bibinfo {author}
  {\bibfnamefont {M.}~\bibnamefont {Tsukada}},\ and\ \bibinfo {author}
  {\bibfnamefont {H.}~\bibnamefont {Fukuyama}},\ }\href
  {https://doi.org/https://doi.org/10.1016/j.apsusc.2004.09.091} {\bibfield
  {journal} {\bibinfo  {journal} {Applied Surface Science}\ }\textbf {\bibinfo
  {volume} {241}},\ \bibinfo {pages} {43 } (\bibinfo {year} {2005})},\ \bibinfo
  {note} {the 9th International Symposium on Advanced Physical
  Fields}\BibitemShut {NoStop}%
\bibitem [{\citenamefont {Sugawara}\ \emph {et~al.}(2006)\citenamefont
  {Sugawara}, \citenamefont {Sato}, \citenamefont {Souma}, \citenamefont
  {Takahashi},\ and\ \citenamefont {Suematsu}}]{PhysRevB.73.045124}%
  \BibitemOpen
  \bibfield  {author} {\bibinfo {author} {\bibfnamefont {K.}~\bibnamefont
  {Sugawara}}, \bibinfo {author} {\bibfnamefont {T.}~\bibnamefont {Sato}},
  \bibinfo {author} {\bibfnamefont {S.}~\bibnamefont {Souma}}, \bibinfo
  {author} {\bibfnamefont {T.}~\bibnamefont {Takahashi}},\ and\ \bibinfo
  {author} {\bibfnamefont {H.}~\bibnamefont {Suematsu}},\ }\href
  {https://doi.org/10.1103/PhysRevB.73.045124} {\bibfield  {journal} {\bibinfo
  {journal} {Phys. Rev. B}\ }\textbf {\bibinfo {volume} {73}},\ \bibinfo
  {pages} {045124} (\bibinfo {year} {2006})}\BibitemShut {NoStop}%
\bibitem [{\citenamefont {Ziatdinov}\ \emph {et~al.}(2013)\citenamefont
  {Ziatdinov}, \citenamefont {Fujii}, \citenamefont {Kusakabe}, \citenamefont
  {Kiguchi}, \citenamefont {Mori},\ and\ \citenamefont
  {Enoki}}]{PhysRevB.87.115427}%
  \BibitemOpen
  \bibfield  {author} {\bibinfo {author} {\bibfnamefont {M.}~\bibnamefont
  {Ziatdinov}}, \bibinfo {author} {\bibfnamefont {S.}~\bibnamefont {Fujii}},
  \bibinfo {author} {\bibfnamefont {K.}~\bibnamefont {Kusakabe}}, \bibinfo
  {author} {\bibfnamefont {M.}~\bibnamefont {Kiguchi}}, \bibinfo {author}
  {\bibfnamefont {T.}~\bibnamefont {Mori}},\ and\ \bibinfo {author}
  {\bibfnamefont {T.}~\bibnamefont {Enoki}},\ }\href
  {https://doi.org/10.1103/PhysRevB.87.115427} {\bibfield  {journal} {\bibinfo
  {journal} {Phys. Rev. B}\ }\textbf {\bibinfo {volume} {87}},\ \bibinfo
  {pages} {115427} (\bibinfo {year} {2013})}\BibitemShut {NoStop}%
\bibitem [{\citenamefont {Fujii}\ \emph {et~al.}(2014)\citenamefont {Fujii},
  \citenamefont {Ziatdinov}, \citenamefont {Ohtsuka}, \citenamefont {Kusakabe},
  \citenamefont {Kiguchi},\ and\ \citenamefont {Enoki}}]{Fujii2014}%
  \BibitemOpen
  \bibfield  {author} {\bibinfo {author} {\bibfnamefont {S.}~\bibnamefont
  {Fujii}}, \bibinfo {author} {\bibfnamefont {M.}~\bibnamefont {Ziatdinov}},
  \bibinfo {author} {\bibfnamefont {M.}~\bibnamefont {Ohtsuka}}, \bibinfo
  {author} {\bibfnamefont {K.}~\bibnamefont {Kusakabe}}, \bibinfo {author}
  {\bibfnamefont {M.}~\bibnamefont {Kiguchi}},\ and\ \bibinfo {author}
  {\bibfnamefont {T.}~\bibnamefont {Enoki}},\ }\href
  {https://doi.org/10.1039/c4fd00073k} {\bibfield  {journal} {\bibinfo
  {journal} {Faraday Discuss.}\ }\textbf {\bibinfo {volume} {173}},\ \bibinfo
  {pages} {173} (\bibinfo {year} {2014})}\BibitemShut {NoStop}%
\bibitem [{\citenamefont {Ziatdinov}\ \emph {et~al.}(2017)\citenamefont
  {Ziatdinov}, \citenamefont {Lim}, \citenamefont {Fujii}, \citenamefont
  {Kusakabe}, \citenamefont {Kiguchi}, \citenamefont {Enoki},\ and\
  \citenamefont {Kim}}]{Ziatdinov2017}%
  \BibitemOpen
  \bibfield  {author} {\bibinfo {author} {\bibfnamefont {M.}~\bibnamefont
  {Ziatdinov}}, \bibinfo {author} {\bibfnamefont {H.}~\bibnamefont {Lim}},
  \bibinfo {author} {\bibfnamefont {S.}~\bibnamefont {Fujii}}, \bibinfo
  {author} {\bibfnamefont {K.}~\bibnamefont {Kusakabe}}, \bibinfo {author}
  {\bibfnamefont {M.}~\bibnamefont {Kiguchi}}, \bibinfo {author} {\bibfnamefont
  {T.}~\bibnamefont {Enoki}},\ and\ \bibinfo {author} {\bibfnamefont
  {Y.}~\bibnamefont {Kim}},\ }\href {https://doi.org/10.1039/c6cp08352h}
  {\bibfield  {journal} {\bibinfo  {journal} {Physical Chemistry Chemical
  Physics}\ }\textbf {\bibinfo {volume} {19}},\ \bibinfo {pages} {5145}
  (\bibinfo {year} {2017})}\BibitemShut {NoStop}%
\bibitem [{\citenamefont {Kusakabe}\ \emph {et~al.}(2018)\citenamefont
  {Kusakabe}, \citenamefont {Nishiguchi}, \citenamefont {Teranishi},\ and\
  \citenamefont {Wicaksono}}]{doi:10.7566/JPSJ.87.084706}%
  \BibitemOpen
  \bibfield  {author} {\bibinfo {author} {\bibfnamefont {K.}~\bibnamefont
  {Kusakabe}}, \bibinfo {author} {\bibfnamefont {K.}~\bibnamefont
  {Nishiguchi}}, \bibinfo {author} {\bibfnamefont {S.}~\bibnamefont
  {Teranishi}},\ and\ \bibinfo {author} {\bibfnamefont {Y.}~\bibnamefont
  {Wicaksono}},\ }\href {https://doi.org/10.7566/JPSJ.87.084706} {\bibfield
  {journal} {\bibinfo  {journal} {Journal of the Physical Society of Japan}\
  }\textbf {\bibinfo {volume} {87}},\ \bibinfo {pages} {084706} (\bibinfo
  {year} {2018})}\BibitemShut {NoStop}%
\bibitem [{\citenamefont {Pereira}\ \emph {et~al.}(2006)\citenamefont
  {Pereira}, \citenamefont {Guinea}, \citenamefont {Lopes~dos Santos},
  \citenamefont {Peres},\ and\ \citenamefont
  {Castro~Neto}}]{PhysRevLett.96.036801}%
  \BibitemOpen
  \bibfield  {author} {\bibinfo {author} {\bibfnamefont {V.~M.}\ \bibnamefont
  {Pereira}}, \bibinfo {author} {\bibfnamefont {F.}~\bibnamefont {Guinea}},
  \bibinfo {author} {\bibfnamefont {J.~M.~B.}\ \bibnamefont {Lopes~dos
  Santos}}, \bibinfo {author} {\bibfnamefont {N.~M.~R.}\ \bibnamefont
  {Peres}},\ and\ \bibinfo {author} {\bibfnamefont {A.~H.}\ \bibnamefont
  {Castro~Neto}},\ }\href {https://doi.org/10.1103/PhysRevLett.96.036801}
  {\bibfield  {journal} {\bibinfo  {journal} {Phys. Rev. Lett.}\ }\textbf
  {\bibinfo {volume} {96}},\ \bibinfo {pages} {036801} (\bibinfo {year}
  {2006})}\BibitemShut {NoStop}%
\bibitem [{\citenamefont {Pereira}\ \emph {et~al.}(2007)\citenamefont
  {Pereira}, \citenamefont {Guinea}, \citenamefont {Lopes~dos Santos},
  \citenamefont {Peres},\ and\ \citenamefont
  {Castro~Neto}}]{PhysRevLett.98.259902}%
  \BibitemOpen
  \bibfield  {author} {\bibinfo {author} {\bibfnamefont {V.~M.}\ \bibnamefont
  {Pereira}}, \bibinfo {author} {\bibfnamefont {F.}~\bibnamefont {Guinea}},
  \bibinfo {author} {\bibfnamefont {J.~M.~B.}\ \bibnamefont {Lopes~dos
  Santos}}, \bibinfo {author} {\bibfnamefont {N.~M.~R.}\ \bibnamefont
  {Peres}},\ and\ \bibinfo {author} {\bibfnamefont {A.~H.}\ \bibnamefont
  {Castro~Neto}},\ }\href {https://doi.org/10.1103/PhysRevLett.98.259902}
  {\bibfield  {journal} {\bibinfo  {journal} {Phys. Rev. Lett.}\ }\textbf
  {\bibinfo {volume} {98}},\ \bibinfo {pages} {259902} (\bibinfo {year}
  {2007})}\BibitemShut {NoStop}%
\bibitem [{\citenamefont {Pereira}\ \emph {et~al.}(2008)\citenamefont
  {Pereira}, \citenamefont {Lopes~dos Santos},\ and\ \citenamefont
  {Castro~Neto}}]{PhysRevB.77.115109}%
  \BibitemOpen
  \bibfield  {author} {\bibinfo {author} {\bibfnamefont {V.~M.}\ \bibnamefont
  {Pereira}}, \bibinfo {author} {\bibfnamefont {J.~M.~B.}\ \bibnamefont
  {Lopes~dos Santos}},\ and\ \bibinfo {author} {\bibfnamefont {A.~H.}\
  \bibnamefont {Castro~Neto}},\ }\href
  {https://doi.org/10.1103/PhysRevB.77.115109} {\bibfield  {journal} {\bibinfo
  {journal} {Phys. Rev. B}\ }\textbf {\bibinfo {volume} {77}},\ \bibinfo
  {pages} {115109} (\bibinfo {year} {2008})}\BibitemShut {NoStop}%
\bibitem [{\citenamefont {Lehtinen}\ \emph {et~al.}(2004)\citenamefont
  {Lehtinen}, \citenamefont {Foster}, \citenamefont {Ma}, \citenamefont
  {Krasheninnikov},\ and\ \citenamefont {Nieminen}}]{PhysRevLett.93.187202}%
  \BibitemOpen
  \bibfield  {author} {\bibinfo {author} {\bibfnamefont {P.~O.}\ \bibnamefont
  {Lehtinen}}, \bibinfo {author} {\bibfnamefont {A.~S.}\ \bibnamefont
  {Foster}}, \bibinfo {author} {\bibfnamefont {Y.}~\bibnamefont {Ma}}, \bibinfo
  {author} {\bibfnamefont {A.~V.}\ \bibnamefont {Krasheninnikov}},\ and\
  \bibinfo {author} {\bibfnamefont {R.~M.}\ \bibnamefont {Nieminen}},\ }\href
  {https://doi.org/10.1103/PhysRevLett.93.187202} {\bibfield  {journal}
  {\bibinfo  {journal} {Phys. Rev. Lett.}\ }\textbf {\bibinfo {volume} {93}},\
  \bibinfo {pages} {187202} (\bibinfo {year} {2004})}\BibitemShut {NoStop}%
\bibitem [{\citenamefont {Ziatdinov}\ \emph {et~al.}(2014)\citenamefont
  {Ziatdinov}, \citenamefont {Fujii}, \citenamefont {Kusakabe}, \citenamefont
  {Kiguchi}, \citenamefont {Mori},\ and\ \citenamefont
  {Enoki}}]{PhysRevB.89.155405}%
  \BibitemOpen
  \bibfield  {author} {\bibinfo {author} {\bibfnamefont {M.}~\bibnamefont
  {Ziatdinov}}, \bibinfo {author} {\bibfnamefont {S.}~\bibnamefont {Fujii}},
  \bibinfo {author} {\bibfnamefont {K.}~\bibnamefont {Kusakabe}}, \bibinfo
  {author} {\bibfnamefont {M.}~\bibnamefont {Kiguchi}}, \bibinfo {author}
  {\bibfnamefont {T.}~\bibnamefont {Mori}},\ and\ \bibinfo {author}
  {\bibfnamefont {T.}~\bibnamefont {Enoki}},\ }\href
  {https://doi.org/10.1103/PhysRevB.89.155405} {\bibfield  {journal} {\bibinfo
  {journal} {Phys. Rev. B}\ }\textbf {\bibinfo {volume} {89}},\ \bibinfo
  {pages} {155405} (\bibinfo {year} {2014})}\BibitemShut {NoStop}%
\bibitem [{\citenamefont {Aoki}\ and\ \citenamefont
  {Dresselhaus}(2014)}]{2014}%
  \BibitemOpen
  \bibinfo {editor} {\bibfnamefont {H.}~\bibnamefont {Aoki}}\ and\ \bibinfo
  {editor} {\bibfnamefont {M.~S.}\ \bibnamefont {Dresselhaus}},\ eds.,\ \href
  {https://doi.org/10.1007/978-3-319-02633-6} {\emph {\bibinfo {title} {Physics
  of Graphene}}}\ (\bibinfo  {publisher} {Springer International Publishing},\
  \bibinfo {year} {2014})\BibitemShut {NoStop}%
\bibitem [{\citenamefont {Enoki}\ and\ \citenamefont {Ando}(2019)}]{2019}%
  \BibitemOpen
  \bibinfo {editor} {\bibfnamefont {T.}~\bibnamefont {Enoki}}\ and\ \bibinfo
  {editor} {\bibfnamefont {T.}~\bibnamefont {Ando}},\ eds.,\ \href
  {https://www.jennystanford.com/9789814800389/physics-and-chemistry-of-graphene-second-edition/}
  {\emph {\bibinfo {title} {Physics and Chemistry of Graphene (Second
  Edition)}}}\ (\bibinfo  {publisher} {Jenny Stanford Publishing},\ \bibinfo
  {year} {2019})\BibitemShut {NoStop}%
\bibitem [{\citenamefont {Shima}\ and\ \citenamefont
  {Aoki}(1993)}]{PhysRevLett.71.4389}%
  \BibitemOpen
  \bibfield  {author} {\bibinfo {author} {\bibfnamefont {N.}~\bibnamefont
  {Shima}}\ and\ \bibinfo {author} {\bibfnamefont {H.}~\bibnamefont {Aoki}},\
  }\href {https://doi.org/10.1103/PhysRevLett.71.4389} {\bibfield  {journal}
  {\bibinfo  {journal} {Phys. Rev. Lett.}\ }\textbf {\bibinfo {volume} {71}},\
  \bibinfo {pages} {4389} (\bibinfo {year} {1993})}\BibitemShut {NoStop}%
\bibitem [{\citenamefont {Ferreira}\ and\ \citenamefont
  {Mucciolo}(2015)}]{PhysRevLett.115.106601}%
  \BibitemOpen
  \bibfield  {author} {\bibinfo {author} {\bibfnamefont {A.}~\bibnamefont
  {Ferreira}}\ and\ \bibinfo {author} {\bibfnamefont {E.~R.}\ \bibnamefont
  {Mucciolo}},\ }\href {https://doi.org/10.1103/PhysRevLett.115.106601}
  {\bibfield  {journal} {\bibinfo  {journal} {Phys. Rev. Lett.}\ }\textbf
  {\bibinfo {volume} {115}},\ \bibinfo {pages} {106601} (\bibinfo {year}
  {2015})}\BibitemShut {NoStop}%
\bibitem [{\citenamefont {Kot}\ \emph {et~al.}(2020)\citenamefont {Kot},
  \citenamefont {Parnell}, \citenamefont {Habibian}, \citenamefont
  {Stra\ss{}er}, \citenamefont {Ostrovsky},\ and\ \citenamefont
  {Ast}}]{PhysRevB.101.235116}%
  \BibitemOpen
  \bibfield  {author} {\bibinfo {author} {\bibfnamefont {P.}~\bibnamefont
  {Kot}}, \bibinfo {author} {\bibfnamefont {J.}~\bibnamefont {Parnell}},
  \bibinfo {author} {\bibfnamefont {S.}~\bibnamefont {Habibian}}, \bibinfo
  {author} {\bibfnamefont {C.}~\bibnamefont {Stra\ss{}er}}, \bibinfo {author}
  {\bibfnamefont {P.~M.}\ \bibnamefont {Ostrovsky}},\ and\ \bibinfo {author}
  {\bibfnamefont {C.~R.}\ \bibnamefont {Ast}},\ }\href
  {https://doi.org/10.1103/PhysRevB.101.235116} {\bibfield  {journal} {\bibinfo
   {journal} {Phys. Rev. B}\ }\textbf {\bibinfo {volume} {101}},\ \bibinfo
  {pages} {235116} (\bibinfo {year} {2020})}\BibitemShut {NoStop}%
\bibitem [{\citenamefont {Morishita}\ \emph {et~al.}(2016)\citenamefont
  {Morishita}, \citenamefont {Sunnardianto}, \citenamefont {Miyao},\ and\
  \citenamefont {Kusakabe}}]{doi:10.7566/JPSJ.85.084703}%
  \BibitemOpen
  \bibfield  {author} {\bibinfo {author} {\bibfnamefont {N.}~\bibnamefont
  {Morishita}}, \bibinfo {author} {\bibfnamefont {G.~K.}\ \bibnamefont
  {Sunnardianto}}, \bibinfo {author} {\bibfnamefont {S.}~\bibnamefont
  {Miyao}},\ and\ \bibinfo {author} {\bibfnamefont {K.}~\bibnamefont
  {Kusakabe}},\ }\href {https://doi.org/10.7566/JPSJ.85.084703} {\bibfield
  {journal} {\bibinfo  {journal} {Journal of the Physical Society of Japan}\
  }\textbf {\bibinfo {volume} {85}},\ \bibinfo {pages} {084703} (\bibinfo
  {year} {2016})}\BibitemShut {NoStop}%
\bibitem [{\citenamefont {Miyao}\ \emph {et~al.}(2017)\citenamefont {Miyao},
  \citenamefont {Morishita}, \citenamefont {Sunnardianto},\ and\ \citenamefont
  {Kusakabe}}]{doi:10.7566/JPSJ.86.034802}%
  \BibitemOpen
  \bibfield  {author} {\bibinfo {author} {\bibfnamefont {S.}~\bibnamefont
  {Miyao}}, \bibinfo {author} {\bibfnamefont {N.}~\bibnamefont {Morishita}},
  \bibinfo {author} {\bibfnamefont {G.~K.}\ \bibnamefont {Sunnardianto}},\ and\
  \bibinfo {author} {\bibfnamefont {K.}~\bibnamefont {Kusakabe}},\ }\href
  {https://doi.org/10.7566/JPSJ.86.034802} {\bibfield  {journal} {\bibinfo
  {journal} {Journal of the Physical Society of Japan}\ }\textbf {\bibinfo
  {volume} {86}},\ \bibinfo {pages} {034802} (\bibinfo {year}
  {2017})}\BibitemShut {NoStop}%
\bibitem [{\citenamefont {Weik}\ \emph {et~al.}(2016)\citenamefont {Weik},
  \citenamefont {Schindler}, \citenamefont {Bera}, \citenamefont {Solomon},\
  and\ \citenamefont {Evers}}]{PhysRevB.94.064204}%
  \BibitemOpen
  \bibfield  {author} {\bibinfo {author} {\bibfnamefont {N.}~\bibnamefont
  {Weik}}, \bibinfo {author} {\bibfnamefont {J.}~\bibnamefont {Schindler}},
  \bibinfo {author} {\bibfnamefont {S.}~\bibnamefont {Bera}}, \bibinfo {author}
  {\bibfnamefont {G.~C.}\ \bibnamefont {Solomon}},\ and\ \bibinfo {author}
  {\bibfnamefont {F.}~\bibnamefont {Evers}},\ }\href
  {https://doi.org/10.1103/PhysRevB.94.064204} {\bibfield  {journal} {\bibinfo
  {journal} {Phys. Rev. B}\ }\textbf {\bibinfo {volume} {94}},\ \bibinfo
  {pages} {064204} (\bibinfo {year} {2016})}\BibitemShut {NoStop}%
\bibitem [{\citenamefont {Ovdat}\ \emph {et~al.}(2020)\citenamefont {Ovdat},
  \citenamefont {Don},\ and\ \citenamefont {Akkermans}}]{PhysRevB.102.075109}%
  \BibitemOpen
  \bibfield  {author} {\bibinfo {author} {\bibfnamefont {O.}~\bibnamefont
  {Ovdat}}, \bibinfo {author} {\bibfnamefont {Y.}~\bibnamefont {Don}},\ and\
  \bibinfo {author} {\bibfnamefont {E.}~\bibnamefont {Akkermans}},\ }\href
  {https://doi.org/10.1103/PhysRevB.102.075109} {\bibfield  {journal} {\bibinfo
   {journal} {Phys. Rev. B}\ }\textbf {\bibinfo {volume} {102}},\ \bibinfo
  {pages} {075109} (\bibinfo {year} {2020})}\BibitemShut {NoStop}%
\bibitem [{\citenamefont {Morishita}\ and\ \citenamefont
  {Kusakabe}(2019)}]{doi:10.7566/JPSJ.88.124707}%
  \BibitemOpen
  \bibfield  {author} {\bibinfo {author} {\bibfnamefont {N.}~\bibnamefont
  {Morishita}}\ and\ \bibinfo {author} {\bibfnamefont {K.}~\bibnamefont
  {Kusakabe}},\ }\href {https://doi.org/10.7566/JPSJ.88.124707} {\bibfield
  {journal} {\bibinfo  {journal} {Journal of the Physical Society of Japan}\
  }\textbf {\bibinfo {volume} {88}},\ \bibinfo {pages} {124707} (\bibinfo
  {year} {2019})}\BibitemShut {NoStop}%
\bibitem [{\citenamefont {Ovchinnikov}(1978)}]{Ovchinnikov1978}%
  \BibitemOpen
  \bibfield  {author} {\bibinfo {author} {\bibfnamefont {A.~A.}\ \bibnamefont
  {Ovchinnikov}},\ }\href {https://doi.org/10.1007/bf00549259} {\bibfield
  {journal} {\bibinfo  {journal} {Theoretica Chimica Acta}\ }\textbf {\bibinfo
  {volume} {47}},\ \bibinfo {pages} {297} (\bibinfo {year} {1978})}\BibitemShut
  {NoStop}%
\bibitem [{\citenamefont {Lieb}(1989)}]{PhysRevLett.62.1201}%
  \BibitemOpen
  \bibfield  {author} {\bibinfo {author} {\bibfnamefont {E.~H.}\ \bibnamefont
  {Lieb}},\ }\href {https://doi.org/10.1103/PhysRevLett.62.1201} {\bibfield
  {journal} {\bibinfo  {journal} {Phys. Rev. Lett.}\ }\textbf {\bibinfo
  {volume} {62}},\ \bibinfo {pages} {1201} (\bibinfo {year}
  {1989})}\BibitemShut {NoStop}%
\bibitem [{\citenamefont {Ortiz}\ \emph {et~al.}(2019)\citenamefont {Ortiz},
  \citenamefont {Boto}, \citenamefont {Garc{\'{\i}}a-Mart{\'{\i}}nez},
  \citenamefont {Sancho-Garc{\'{\i}}a}, \citenamefont {Melle-Franco},\ and\
  \citenamefont {Fern{\'{a}}ndez-Rossier}}]{doi:10.1021/acs.nanolett.9b01773}%
  \BibitemOpen
  \bibfield  {author} {\bibinfo {author} {\bibfnamefont {R.}~\bibnamefont
  {Ortiz}}, \bibinfo {author} {\bibfnamefont {R.~A.}\ \bibnamefont {Boto}},
  \bibinfo {author} {\bibfnamefont {N.}~\bibnamefont
  {Garc{\'{\i}}a-Mart{\'{\i}}nez}}, \bibinfo {author} {\bibfnamefont {J.~C.}\
  \bibnamefont {Sancho-Garc{\'{\i}}a}}, \bibinfo {author} {\bibfnamefont
  {M.}~\bibnamefont {Melle-Franco}},\ and\ \bibinfo {author} {\bibfnamefont
  {J.}~\bibnamefont {Fern{\'{a}}ndez-Rossier}},\ }\href
  {https://doi.org/10.1021/acs.nanolett.9b01773} {\bibfield  {journal}
  {\bibinfo  {journal} {Nano Letters}\ }\textbf {\bibinfo {volume} {19}},\
  \bibinfo {pages} {5991} (\bibinfo {year} {2019})}\BibitemShut {NoStop}%
\bibitem [{\citenamefont {Gil-Guerrero}\ \emph {et~al.}(2020)\citenamefont
  {Gil-Guerrero}, \citenamefont {Melle-Franco}, \citenamefont
  {Pe{\~{n}}a-Gallego},\ and\ \citenamefont
  {Mandado}}]{https://doi.org/10.1002/chem.202003713}%
  \BibitemOpen
  \bibfield  {author} {\bibinfo {author} {\bibfnamefont {S.}~\bibnamefont
  {Gil-Guerrero}}, \bibinfo {author} {\bibfnamefont {M.}~\bibnamefont
  {Melle-Franco}}, \bibinfo {author} {\bibfnamefont {{\'{A}}.}~\bibnamefont
  {Pe{\~{n}}a-Gallego}},\ and\ \bibinfo {author} {\bibfnamefont
  {M.}~\bibnamefont {Mandado}},\ }\href
  {https://doi.org/10.1002/chem.202003713} {\bibfield  {journal} {\bibinfo
  {journal} {Chemistry {\textendash} A European Journal}\ }\textbf {\bibinfo
  {volume} {26}},\ \bibinfo {pages} {16138} (\bibinfo {year}
  {2020})}\BibitemShut {NoStop}%
\bibitem [{\citenamefont {Kadanoff}(1966)}]{PhysicsPhysiqueFizika.2.263}%
  \BibitemOpen
  \bibfield  {author} {\bibinfo {author} {\bibfnamefont {L.~P.}\ \bibnamefont
  {Kadanoff}},\ }\href {https://doi.org/10.1103/PhysicsPhysiqueFizika.2.263}
  {\bibfield  {journal} {\bibinfo  {journal} {Physics Physique Fizika}\
  }\textbf {\bibinfo {volume} {2}},\ \bibinfo {pages} {263} (\bibinfo {year}
  {1966})}\BibitemShut {NoStop}%
\bibitem [{\citenamefont {Kadanoff}\ \emph {et~al.}(1967)\citenamefont
  {Kadanoff}, \citenamefont {G\"otze}, \citenamefont {Hamblen}, \citenamefont
  {Hecht}, \citenamefont {Lewis}, \citenamefont {Palciauskas}, \citenamefont
  {Rayl}, \citenamefont {Swift}, \citenamefont {Aspens},\ and\ \citenamefont
  {Kane}}]{RevModPhys.39.395}%
  \BibitemOpen
  \bibfield  {author} {\bibinfo {author} {\bibfnamefont {L.~P.}\ \bibnamefont
  {Kadanoff}}, \bibinfo {author} {\bibfnamefont {W.}~\bibnamefont {G\"otze}},
  \bibinfo {author} {\bibfnamefont {D.}~\bibnamefont {Hamblen}}, \bibinfo
  {author} {\bibfnamefont {R.}~\bibnamefont {Hecht}}, \bibinfo {author}
  {\bibfnamefont {E.~A.~S.}\ \bibnamefont {Lewis}}, \bibinfo {author}
  {\bibfnamefont {V.~V.}\ \bibnamefont {Palciauskas}}, \bibinfo {author}
  {\bibfnamefont {M.}~\bibnamefont {Rayl}}, \bibinfo {author} {\bibfnamefont
  {J.}~\bibnamefont {Swift}}, \bibinfo {author} {\bibfnamefont
  {D.}~\bibnamefont {Aspens}},\ and\ \bibinfo {author} {\bibfnamefont
  {J.}~\bibnamefont {Kane}},\ }\href
  {https://doi.org/10.1103/RevModPhys.39.395} {\bibfield  {journal} {\bibinfo
  {journal} {Rev. Mod. Phys.}\ }\textbf {\bibinfo {volume} {39}},\ \bibinfo
  {pages} {395} (\bibinfo {year} {1967})}\BibitemShut {NoStop}%
\end{thebibliography}
%


\end{document}